\documentclass[11pt,a4paper]{article}
\pdfoutput=1
\usepackage[utf8]{inputenc}
\usepackage{a4wide}
\usepackage{amsmath}
\usepackage{amssymb}
\usepackage{amsfonts}
\usepackage{cite}
\usepackage{color}
\usepackage{adjustbox}
\usepackage{multirow}
\usepackage{pdflscape}
\usepackage{textcomp}
\usepackage{gensymb}
\usepackage{graphicx}
\usepackage{diagbox}
\usepackage{pifont}
\usepackage{bigstrut}
\usepackage[small,bf]{caption}
\usepackage{tabulary}
\usepackage{booktabs}
\usepackage{stackengine}
\usepackage{mathtools}
\usepackage{enumerate}
\usepackage{makecell}
\usepackage{listings}
\usepackage[unicode]{hyperref}

\newcommand{\be}{\begin{equation}}
\newcommand{\ee}{\end{equation}}

\def\1{\mathbf{1}}

\def\2{\mathbf{2}}
\def\3{\mathbf{3}}
\def\tp{\mathbf{3^\prime}}
\def\5{\mathbf{5}}

\def\g{\gamma}
\def\G{\Gamma}

\def\L{\Lambda}

\def\r{\rho}

\def\t{\tau}
\def\th{\theta}

\def\ot{\otimes}

\def\di{{\rm d}}

\DeclareMathOperator{\diag}{diag}

\DeclareMathOperator{\im}{Im}
\DeclareMathOperator{\re}{Re}

\DeclarePairedDelimiter{\vevdel}{\langle}{\rangle}
\newcommand{\vev}{\vevdel}
\allowdisplaybreaks
\numberwithin{equation}{section}
\makeatletter
\g@addto@macro\bfseries{\boldmath}
\makeatother

\begin{document}

\begin{titlepage}

\vspace*{-15mm}
\begin{flushright}
SISSA  54/2018/FISI \\
IPMU18-0202 \\
IPPP/18/105
\end{flushright}
\vspace*{8mm}

\begin{center}
{\bf\LARGE Modular $A_5$ Symmetry for Flavour Model Building} \\[8mm]
P.~P.~Novichkov$^{\,a,}$\footnote{E-mail: \texttt{pavel.novichkov@sissa.it}}, 
J.~T.~Penedo$^{\,b,}$\footnote{E-mail: \texttt{joao.t.n.penedo@tecnico.ulisboa.pt}}, 
S.~T.~Petcov$^{\,a,c,}$\footnote{Also at
Institute of Nuclear Research and Nuclear Energy,
Bulgarian Academy of Sciences, 1784 Sofia, Bulgaria.},
A.~V.~Titov$^{\,d,}$\footnote{E-mail: \texttt{arsenii.titov@durham.ac.uk}}\\
\vspace{8mm}
$^{a}$\,{\it SISSA/INFN, Via Bonomea 265, 34136 Trieste, Italy} \\
\vspace{2mm}
$^{b}$\,{\it CFTP, Departamento de Física, Instituto Superior Técnico, Universidade de Lisboa,\\
Avenida Rovisco Pais 1, 1049-001 Lisboa, Portugal} \\
\vspace{2mm}
$^{c}$\,{\it Kavli IPMU (WPI), University of Tokyo, 5-1-5 Kashiwanoha, 277-8583 Kashiwa, Japan} \\
\vspace{2mm}
$^{d}$\,{\it Institute for Particle Physics Phenomenology, 
Department of Physics, Durham University,\\ 
South Road, Durham DH1 3LE, United Kingdom}
\end{center}
\vspace{8mm}

\begin{abstract}
\noindent
In the framework of the modular symmetry approach 
to lepton flavour, we consider a class of theories 
where matter superfields transform in representations of
the finite modular group $\G_5 \simeq A_5$. 
We explicitly construct a basis for 
the 11 modular forms of weight 2 and level 5. 
We show how these forms arrange themselves 
into two triplets and a quintet of~$A_5$. 
We also present multiplets of modular forms 
of higher weight. 
Finally, we provide an example of
application of our results, constructing
two models of 
neutrino masses and mixing 
based on the supersymmetric Weinberg operator.
\end{abstract}

\end{titlepage}
\setcounter{footnote}{0}

\section{Introduction}
\label{sec:intro}
Understanding the origins of flavour remains one of 
the major problems in particle physics. 
The power of symmetries in governing laws 
of particle interactions does not need to be advocated. 
In this regard, it is rather natural to expect 
that symmetry(ies) also hold the key to the solution 
of the flavour problem. 

 The fact that two out of three neutrino mixing angles are large 
\cite{PDG2018,Capozzi:2018ubv,Esteban:2018azc} suggests the presence of a new flavour symmetry 
(at least in the lepton sector)
described by 
a non-Abelian discrete (finite) group (see, e.g., 
\cite{Altarelli:2010gt,Ishimori:2010au,King:2013eh,Petcov:2017ggy}). 
While unifying the three known flavours at high energies, 
this symmetry may be broken at lower energies to residual symmetries 
of the charged lepton and neutrino mass terms, 
which correspond to Abelian subgroups of the original flavour symmetry group.
In the bottom-up approach, starting from residual symmetries, 
one can successfully explain the observed pattern of neutrino mixing 
and, in addition, predict the value of the Dirac CP violation phase 
\cite{Fonseca:2014koa,Girardi:2015rwa,Petcov:2018snn}.%
\footnote{Predictions for the Dirac CPV phase 
can be obtained also if the neutrino Majorana mass matrix 
respects a specific residual symmetry while the mixing originating 
from the charged lepton sector has a form restricted by 
additional (GUT, generalised CP) symmetry or phenomenological 
considerations, see, e.g., 
\cite{Girardi:2013sza,Petcov:2014laa,Girardi:2015vha}.}
However, predicting neutrino masses calls for the  construction of specific models, 
in which the flavour symmetry is typically spontaneously broken 
by vacuum expectations values (VEVs) of flavons~---~scalar gauge singlets 
charged non-trivially under the flavour symmetry group. 
Usually, a numerous set of these fields is needed.
Moreover, one may have to construct rather complicated flavon potentials
in order to achieve vacuum alignments leading 
to viable phenomenology.

  A very interesting generalisation of 
the discrete symmetry approach to lepton flavour
has been recently proposed in Ref.~\cite{Feruglio:2017spp}.
In this proposal, modular invariance plays the role of flavour symmetry, 
and couplings of a theory are modular forms of a certain level $N$.
In addition, both the couplings and matter supermultiplets are assumed to
transform in representations of a finite modular group $\G_N$. 
In the simplest class of such models, the VEV of a complex field $\t$
(the modulus) is the only source of flavour symmetry breaking, 
such that no flavons are needed. 
Another appealing feature of the proposed framework 
is that charged lepton and neutrino masses, 
neutrino mixing and CPV phases
are simultaneously determined  by the modular symmetry 
typically in terms of a limited number of constant parameters. 
This leads to experimentally testable correlations between, e.g., the neutrino mass and mixing observables.

 The cornerstone of the new approach is 
the modular forms of weight 2 and level $N$, 
and their arrangements into multiplets of $\G_N$. 
Modular forms of higher weights can be constructed from these building blocks.
Remarkably, for $N \leq 5$, the finite modular groups 
are isomorphic to well-known permutation groups.
In Ref.~\cite{Feruglio:2017spp}, the group $\G_3 \simeq A_4$ 
has been considered, and the three generating modular forms of weight 2 
have been explicitly constructed and shown to furnish 
a 3-dimensional irreducible representation (irrep) of $A_4$. 
Further, the group $\G_2 \simeq S_3$ has been considered 
in \cite{Kobayashi:2018vbk}, and the two forms shaping 
a doublet of $S_3$ have been identified. 
The five generating modular forms in the case of $N=4$ 
have been found to organise themselves into 
a doublet and a triplet of $\G_4 \simeq S_4$ in Ref.~\cite{Penedo:2018nmg},
where the  first realistic model of lepton masses and mixing
without flavons has also been constructed.
Very recently, by studying Yukawa couplings 
in magnetised D-brane models, the authors of Ref.~\cite{Kobayashi:2018bff} 
have found multiplets of weight 2 modular forms 
corresponding to a triplet and a sextet of $\Delta(96)$, and a triplet of $\Delta(384)$. 
They have also reported an $S_3$ doublet and an $S_4$ triplet. 
Note that $\Delta(96)$ is isomorphic to a subgroup of $\G_8$, 
while $\Delta(384)$ is isomorphic to a subgroup of $\G_{16}$ 
(see, e.g., \cite{deAdelhartToorop:2011re}).

 Lepton flavour models based on 
$\G_3 \simeq A_4$ have been studied in more detail in
Refs.~\cite{Criado:2018thu,Kobayashi:2018scp}, 
where several viable examples have been presented. 
In Ref.~\cite{Novichkov:2018ovf}, we have constructed 
in a systematic way models based on $\G_4 \simeq S_4$, 
in which light neutrino masses are generated via the type I seesaw mechanism 
and where no flavons are introduced. 
We have shown that models with a relatively small number of free parameters 
can successfully describe data on the charged lepton masses, 
neutrino mass-squared differences and mixing angles. 
Furthermore, we have obtained predictions for the neutrino masses and the Dirac and Majorana CPV phases in the neutrino mixing matrix.
In these models, the value of atmospheric mixing angle $\th_{23}$ is correlated with 
i) the Dirac phase $\delta$, 
ii) the sum of neutrino masses, and
iii) the effective Majorana mass in neutrinoless double beta decay. 

 In the present article, for the first time in this context, 
we consider the finite modular group $\G_5 \simeq A_5$.
Our main focus is on constructing the 11 generating modular forms of weight 2 
and demonstrating how they can be arranged into 
multiplets of $A_5$, namely two triplets and a quintet. 
The group $A_5$ has been investigated in the context of 
the conventional discrete symmetry approach in 
Refs.~\cite{Everett:2008et,Feruglio:2011qq,Ding:2011cm,Cooper:2012bd,Varzielas:2013hga,Gehrlein:2014wda} as well as in combination 
with so-called generalised CP symmetry 
in Refs.~\cite{Li:2015jxa,DiIura:2015kfa,Ballett:2015wia,Turner:2015uta,DiIura:2018fnk}.
The characteristic phenomenological feature of the models 
based on the $A_5$ flavour symmetry 
is the golden ratio prediction for the solar mixing angle, 
$\th_{12} = \arctan(1/\varphi) \approx 32\degree$, with 
$\varphi = (1+\sqrt5)/2$ being the golden ratio, 
which is inside the experimentally allowed $3\sigma$ range 
\cite{Capozzi:2018ubv,Esteban:2018azc}.
An interesting theoretical feature of $A_5$ is that 
it is anomaly-free~\cite{Ishimori:2010au}. 

 The article is organised as follows. 
In Section~\ref{sec:framework}, we first summarise the 
modular symmetry approach to lepton masses and mixing 
proposed in Ref.~\cite{Feruglio:2017spp}, 
and then explicitly construct the two $A_5$ triplets and the $A_5$ quintet 
of modular forms of weight 2. 
Next, in Section~\ref{sec:pheno}, we give an example of 
application of the obtained results 
constructing a phenomenologically viable model of neutrino masses and mixing 
based on the Weinberg operator.
Finally, in Section~\ref{sec:conclusions}, we draw our conclusions.

\section{The Framework}
\label{sec:framework}

\subsection{Modular symmetry and modular-invariant theories}
\label{subsec:modinv}
 The modular group $\overline{\G}$  
is the group of linear fractional transformations $\g$ 
acting on the complex variable $\t$ belonging to 
the upper-half complex plane as follows:
\be
\g\t = \frac{a\t + b}{c\t + d}\,,
\quad
\text{where} 
\quad
a,b,c,d \in \mathbb{Z}
\quad
\text{and}
\quad
ad - bc = 1\,,~~{\rm Im}\tau > 0\,.
\label{eq:linfractransform}
\ee
%
The modular group is isomorphic to the projective special linear group 
$PSL(2,\mathbb{Z})$, and it is
generated by two elements $S$ and $T$ satisfying
%
\be
S^2 = \left(ST\right)^3 = I\,,
\ee
$I$ being the identity element of a group.
Representing $S$ and $T$ as
\be
S = \begin{pmatrix}
0 & 1 \\
-1 & 0
\end{pmatrix}\,,
\qquad
T = \begin{pmatrix} 
1 & 1 \\
0 & 1
\end{pmatrix}\,,
\ee
%
one finds
\be
\t \xrightarrow{S} -\frac{1}{\t}\,, 
\qquad
\t \xrightarrow{T} \t + 1\,.
\ee
%

 The group $SL(2,\mathbb{Z}) = \G(1) \equiv \G$ contains a series of infinite normal subgroups 
$\G(N)$, $N = 1,2,3,\dots$:
\be
\G(N) = \left\{
\begin{pmatrix}
a & b \\
c & d
\end{pmatrix} 
\in SL(2,\mathbb{Z})\,, 
\quad
\begin{pmatrix}
a & b \\
c & d
\end{pmatrix} =
\begin{pmatrix}
1 & 0 \\
0 & 1
\end{pmatrix} 
~~(\text{mod } N)
\right\},
\ee
%
called the principal congruence subgroups.
For $N=1$ and $2$, we introduce the groups $\overline{\G}(N) \equiv \G(N)/\{I,-I\}$ 
(note that $\overline{\G}(1) \equiv \overline{\G}$), 
and for $N > 2$, $\overline{\G}(N) \equiv \G(N)$. 
For each $N$, the associated linear fractional transformations 
of the form in eq.~\eqref{eq:linfractransform} 
are in a one-to-one correspondence with 
the elements of $\overline{\G}(N)$. 
The quotient groups $\G_N \equiv \overline{\G}/\overline{\G}(N)$ 
are called finite modular groups. 
For $N \leq 5$, these groups are isomorphic to 
permutation groups widely used to build flavour models
(see, e.g., \cite{deAdelhartToorop:2011re}). 
Namely, $\G_2 \simeq S_3$, $\G_3 \simeq A_4$, $\G_4 \simeq S_4$ and $\G_5 \simeq A_5$.

 Modular forms of weight $k$ and level $N$ are holomorphic functions $f(\t)$ 
transforming under the action of $\overline{\G}(N)$ in the following way:
\be
f\left(\g\t\right) = \left(c\t + d\right)^k f(\t)\,, 
\quad 
\g \in \overline{\G}(N)\,.
\ee
%
Here $k$ is even and non-negative, and $N$ is natural.
Modular forms of weight $k$ and level $N$ span a linear space
of finite dimension.
There exists a basis in this space such that 
a multiplet of modular forms $f_i(\t)$ 
transforms according to
a unitary representation $\rho$ of the finite group $\G_N$:
\be
f_i\left(\g\t\right) = \left(c\t + d\right)^k \rho\left(\g\right)_{ij} f_j(\t)\,, 
\quad 
\g \in \overline{\G}\,.
\label{eq:vvmodforms}
\ee
%

 In the case of $N=2$, the modular forms of weight 2 
span a two-dimensional linear space. 
In a certain basis the two generating modular forms 
transform in the 2-dimensional irrep 
of $S_3$ \cite{Kobayashi:2018vbk}. For level $N=3$, 
weight 2 modular forms 
arrange themselves in a triplet of $A_4$ \cite{Feruglio:2017spp}. 
In the case of $N=4$, the corresponding linear space has dimension 5, 
and weight 2 modular forms group in a doublet and a triplet of $S_4$ \cite{Penedo:2018nmg}. 
For $N = 5$, there are 11 modular forms of weight 2. 
They are organised in two triplets and a quintet of $A_5$. 
In the next subsection, we will explicitly derive them, 
but before that, let us briefly recall how to construct
supersymmetric modular-invariant theories.

 In the case of $\mathcal{N} = 1$ rigid supersymmetry, 
the matter action $\mathcal{S}$ reads
\be
\mathcal{S} = \int \di^4x\, \di^2\th\, \di^2\overline{\th}~ 
K(\t, \overline{\t}, \chi, \overline{\chi}) + 
\int \di^4x\, \di^2\th~W(\t, \chi) + 
\int \di^4x\, \di^2\overline{\th}~\overline{W}(\overline{\t}, \overline{\chi})\,,
\ee
%
where $K$ is the Kähler potential, $W$ is the superpotential 
and $\chi$ denotes a set of chiral supermultiplets contained in the theory 
apart from the modulus $\t$. 
The $\th$ and $\overline{\th}$ denote Graßmann variables.
The modular group acts on 
$\t$ and supermultiplets $\chi_I$ of a sector $I$ of a theory 
in a certain way~\cite{Ferrara:1989bc,Ferrara:1989qb}.
Assuming, in addition, that the supermultiplets $\chi_I$ transform 
according to a 
representation $\rho_I$ of $\G_N$, 
we have
\be
\begin{cases}
\t \rightarrow \dfrac{a\t + b}{c\t + d}\,, \\[4mm]
\chi_I \rightarrow \left(c\t + d\right)^{-k_I} \rho_I(\g)\, \chi_I\,.
\end{cases}
\label{eq:modtransforms}
\ee
%
Note that $\chi_I$ are not modular forms, 
and the weight $(- k_I)$ can be odd and/or negative.
Requiring invariance of $\mathcal{S}$ under eq.~\eqref{eq:modtransforms}  
leads to
\be
\begin{cases}
W(\t, \chi) \rightarrow W(\t,\chi)\,, \\[4mm]
K(\t, \overline{\t}, \chi, \overline{\chi}) \rightarrow K(\t, \overline{\t}, \chi, \overline{\chi}) 
+ f_K(\t,\chi) + \overline{f_K}(\overline{\t},\overline{\chi})\,,
\end{cases}
\ee
%
where the second line represents a Kähler transformation.
The superpotential can be expanded in powers of $\chi_I$:
\be
W(\t, \chi) = \sum_{n} \sum_{\{I_1,\dots,I_n\}}
\left(Y_{I_1\,\dots\,I_n}(\t)\, \chi_{I_1}\dots\chi_{I_n}\right)_\1\,,
\label{eq:superpotentialGen}
\ee
%
where $\1$ stands for an invariant singlet of $\G_N$.
To ensure invariance of $W$ under the transformations 
specified in eq.~\eqref{eq:modtransforms}, 
the functions $Y_{I_1\,\dots\,I_n}(\t)$ must transform as follows:
\be
Y_{I_1\,\dots\,I_n}(\t) \rightarrow (c\t + d)^{k_{Y}} \rho_{Y}(\g)\, Y_{I_1\,\dots\,I_n}(\t)\,,
\label{eq:vvmf}
\ee
%
where $\rho_{Y}$ is a representation of $\G_N$, and $k_{Y}$ and $\rho_{Y}$ 
are such that
\begin{align}
&k_{Y} = k_{I_1} + \dots + k_{I_n}\,, \\[2mm]
&\rho_{Y} \ot \rho_{I_1} \ot \dots \ot \rho_{I_n} \supset \1\,.
\end{align}
%
Thereby, the functions $Y_{I_1\,\dots\,I_n}(\t)$ form 
a multiplet of weight $k_{Y}$ and level $N$ modular forms 
transforming in the representation $\rho_{Y}$ of $\G_N$
(cf.~eq.~\eqref{eq:vvmodforms}).

\subsection{Generators of modular forms of level \texorpdfstring{$N=5$}{N=5}}
The dimension of the space of modular forms of level $N=5$ and lowest nontrivial weight 2 is 11.
Expansions for a standard basis $\{b_1(\tau),\ldots,b_{11}(\tau)\}$ of this space of functions are given in Appendix~\ref{app:sage}. Modular forms of higher weight can be constructed from homogeneous polynomials in these eleven modular forms.
The action of the discrete quotient group $\Gamma_5$ divides the space of lowest weight modular functions into two triplets 
transforming in irreps
$\mathbf{3}$ and $\mathbf{3'}$ and a quintet 
transforming in 
$\mathbf{5}$ of $\Gamma_5 \simeq A_5$ (see also Section 4.4 of Ref.~\cite{Franc:2016hg}).

As in the cases of 
$\G_3 \simeq A_4$ \cite{Feruglio:2017spp}, 
$\G_2 \simeq S_3$ \cite{Kobayashi:2018vbk} and
$\Gamma_4 \simeq S_4$~\cite{Penedo:2018nmg},
the lowest weight modular functions
correspond to linear combinations of logarithmic derivatives
of some ``seed'' functions $\alpha_{i,j}(\tau)$.
These functions form a set which is
in a certain sense closed under the action of $A_5$.
As can be inferred from the results in Ref.~\cite{Mano:2002dr},
a convenient choice for  $\alpha_{i,j}(\tau)$ 
is given by
the Jacobi theta functions
$\theta_3(z(\tau),t(\tau))$,
and they explicitly read:%
\footnote{For properties of these special functions, see, e.g., Refs.~\cite{Farkas:2001th,Kharchev:2015tv}.
In the notations of Ref.~\cite{Farkas:2001th}, $\theta_3 \equiv \theta\begin{bmatrix}0\\0\end{bmatrix}$.}
\begin{equation}
\begin{aligned}[c]
\alpha_{1,-1}(\tau) \,&\equiv\, \theta_3\left( \frac{\tau+1}{2}, 5\tau\right)
\,, \\
\alpha_{1,0}(\tau) \,&\equiv\, \theta_3\left( \frac{\tau+9}{10}, \frac{\tau}{5}\right)
\,, \\
\alpha_{1,1}(\tau) \,&\equiv\, \theta_3\left( \frac{\tau}{10}, \frac{\tau+1}{5}\right)
\,, \\
\alpha_{1,2}(\tau) \,&\equiv\, \theta_3\left( \frac{\tau+1}{10}, \frac{\tau+2}{5}\right)
\,, \\
\alpha_{1,3}(\tau) \,&\equiv\, \theta_3\left( \frac{\tau+2}{10}, \frac{\tau+3}{5}\right)
\,, \\
\alpha_{1,4}(\tau) \,&\equiv\, \theta_3\left( \frac{\tau+3}{10}, \frac{\tau+4}{5}\right)
\,,
\end{aligned}
\qquad
\begin{aligned}[c]
\alpha_{2,-1}(\tau) \,&\equiv\, e^{2 \pi i \tau / 5} \, 
\theta_3\left( \frac{3\tau+1}{2}, 5\tau\right)
\,, \\
\alpha_{2,0}(\tau) \,&\equiv\, \theta_3\left( \frac{\tau+7}{10}, \frac{\tau}{5}\right)
\,, \\
\alpha_{2,1}(\tau) \,&\equiv\, \theta_3\left( \frac{\tau+8}{10}, \frac{\tau+1}{5}\right)
\,, \\
\alpha_{2,2}(\tau) \,&\equiv\, \theta_3\left( \frac{\tau+9}{10}, \frac{\tau+2}{5}\right)
\,, \\
\alpha_{2,3}(\tau) \,&\equiv\, \theta_3\left( \frac{\tau}{10}, \frac{\tau+3}{5}\right)
\,, \\
\alpha_{2,4}(\tau) \,&\equiv\, \theta_3\left( \frac{\tau+1}{10}, \frac{\tau+4}{5}\right)
\,.
\end{aligned}
\label{eq:seed}
\end{equation}

\begin{figure}
\centering
\includegraphics[width=\textwidth]{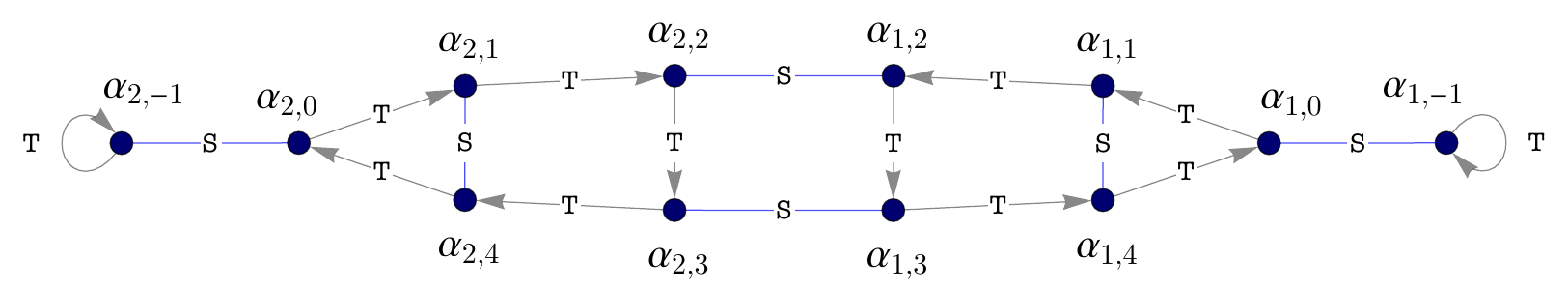}
\caption{Graph illustrating the automorphisms
of the set of seed functions $\alpha_{i,j}(\tau)$,
defined in eq.~\eqref{eq:seed},
under the actions of $\Gamma_5 \simeq A_5$ generators $S$ and $T$.}
\label{fig:graph}
\end{figure}

Under the action of the generators $S$ and $T$ of $\Gamma_5$ 
(see Appendix~\ref{app:basis}), each of these functions is mapped to another, up to (possibly $\tau$-dependent) multiplicative factors.
A diagram of said map is given in Fig.~\ref{fig:graph}, and one can check that
the actions of $S^2$, $(ST)^3$ and $T^5$ applied to each element
correspond to the identity.
Taking logarithmic derivatives, one obtains:
\begin{align}
\frac{\di}{\di\tau}\log\alpha_{i,j}(-1/\tau) &=
\frac{i\pi}{20}\left(1-\frac{1}{\tau^2}\right)+ \frac{1}{2\tau}
+\frac{\di}{\di\tau}\log\alpha^S_{i,j}(\tau)\,,
\\[2mm]
\frac{\di}{\di\tau}\log\alpha_{i,j}(\tau+1) &= 
\frac{\di}{\di\tau}\log\alpha^T_{i,j}(\tau)\,,
\end{align}
where $\alpha^S_{i,j}$ and $\alpha^T_{i,j}$
are the images of $\alpha_{i,j}$
under the $S$ and $T$ maps of Fig.~\ref{fig:graph},
respectively.

It then follows that the functions
\begin{align}
Y(c_{1,-1},\ldots,c_{1,4};c_{2,-1},\ldots,c_{2,4}|\tau) \equiv \sum_{i,j} c_{i,j}
\frac{\di}{\di\tau}\log\alpha_{i,j}(\tau)
\,,\quad \textrm{with } \sum_{i,j} c_{i,j} = 0\,,
\end{align}
span the sought-after 11-dimensional space
of lowest weight modular forms of level $N=5$.
Under $S$ and $T$, one has the following transformations:
\begin{equation}
\begin{aligned}
S:\quad Y(c_{1,-1},\ldots,c_{1,4};c_{2,-1},\ldots,c_{2,4}|\tau)
\,\,\rightarrow\,\,
Y(c_{1,-1},\ldots,c_{1,4};c_{2,-1},\ldots,c_{2,4}|-1/\tau)
\\
= Y(c_{1,0},c_{1,-1},c_{1,4},c_{2,2},c_{2,3},c_{1,1};
  c_{2,0},c_{2,-1},c_{2,4},c_{1,2},c_{1,3},c_{2,1} | \tau)\,,
\end{aligned}
\end{equation}
\begin{equation}
\begin{aligned}
T:\quad Y(c_{1,-1},\ldots,c_{1,4};c_{2,-1},\ldots,c_{2,4}|\tau)
\,\,\rightarrow\,\,
Y(c_{1,-1},\ldots,c_{1,4};c_{2,-1},\ldots,c_{2,4}|\tau+1)
\\
= Y(c_{1,-1},c_{1,4},c_{1,0},c_{1,1},c_{1,2},c_{1,3};
  c_{2,-1},c_{2,4},c_{2,0},c_{2,1},c_{2,2},c_{2,3} | \tau)\,.
\end{aligned}
\end{equation}
Then, as anticipated, the space in question is divided into the following multiplets of $A_5$:
\begin{align}
Y_\mathbf{5}(\tau) = \left(\begin{array}{c}
Y_{1}(\tau)\\
Y_{2}(\tau)\\
Y_{3}(\tau)\\
Y_{4}(\tau)\\
Y_{5}(\tau)
\end{array}\right)
&\equiv\left(\begin{array}{c}
-\frac{1}{\sqrt{6}}Y\left(-5,1,1,1,1,1;-5,1,1,1,1,1\middle|\tau\right)\\
Y(0,1,\zeta^4,\zeta^3,\zeta^2,\zeta\,;\,0,1,\zeta^4,\zeta^3,\zeta^2,\zeta\,|\,\tau)\\
Y(0,1,\zeta^3,\zeta,\zeta^4,\zeta^2\,;\,0,1,\zeta^3,\zeta,\zeta^4,\zeta^2\,|\,\tau)\\
Y(0,1,\zeta^2,\zeta^4,\zeta,\zeta^3\,;\,0,1,\zeta^2,\zeta^4,\zeta,\zeta^3\,|\,\tau)\\
Y(0,1,\zeta,\zeta^2,\zeta^3,\zeta^4\,;\,0,1,\zeta,\zeta^2,\zeta^3,\zeta^4\,|\,\tau)
\end{array}\right)\,,
\label{eq:5} \\[2mm]
Y_\mathbf{3}(\tau) = \left(\begin{array}{c}
Y_{6}(\tau)\\
Y_{7}(\tau)\\
Y_{8}(\tau)
\end{array}\right)
&\equiv\left(\begin{array}{c}
\frac{1}{\sqrt{2}}Y\left(-\sqrt{5},-1,-1,-1,-1,-1;\sqrt{5},1,1,1,1,1\middle|\tau\right)\\
Y(0,1,\zeta^4,\zeta^3,\zeta^2,\zeta\,;\,0,-1,-\zeta^4,-\zeta^3,-\zeta^2,-\zeta\,|\,\tau)\\
Y(0,1,\zeta,\zeta^2,\zeta^3,\zeta^4\,;\,0,-1,-\zeta,-\zeta^2,-\zeta^3,-\zeta^4\,|\,\tau)
\end{array}\right)\,,
\label{eq:3} \\[2mm]
Y_\mathbf{3'}(\tau) = \left(\begin{array}{c}
Y_{9}(\tau)\\
Y_{10}(\tau)\\
Y_{11}(\tau)
\end{array}\right)
&\equiv\left(\begin{array}{c}
\frac{1}{\sqrt{2}}Y\left(\sqrt{5},-1,-1,-1,-1,-1;-\sqrt{5},1,1,1,1,1\middle|\tau\right)\\
Y(0,1,\zeta^3,\zeta,\zeta^4,\zeta^2\,;\,0,-1,-\zeta^3,-\zeta,-\zeta^4,-\zeta^2\,|\,\tau)\\
Y(0,1,\zeta^2,\zeta^4,\zeta,\zeta^3\,;\,0,-1,-\zeta^2,-\zeta^4,-\zeta,-\zeta^3\,|\,\tau)
\end{array}\right)\,,
\label{eq:3p}
\end{align}
where $\zeta = e^{2\pi i/5}$.
The $q$-expansions for these modular forms are given
in Appendix~\ref{app:qexp}.
In Appendix~\ref{app:basis}, we specify our basis choice
for the representation matrices $\rho(\gamma)$ of $A_5$,
and we list the Clebsch-Gordan coefficients for this basis
in Appendix~\ref{app:cbc}.

Multiplets transforming in the other representations of $A_5$
can be obtained from tensor products of the lowest weight multiplets
$Y_\mathbf{5}$, $Y_\mathbf{3}$ and $Y_\mathbf{3'}$.
The missing $\mathbf{1}$ and $\mathbf{4}$ representations
arise at weight 4.
Even though one can form 66 products $Y_i Y_j$,
the dimension of the space of weight 
$k=4$ (and level 5) forms is $5k+1=21$.
Therefore, there are 45 constraints between the $Y_i Y_j$,
which we list in Appendix~\ref{app:constraints},
and which reduce the 66 potentially independent combinations 
to 21 truly independent ones. These last combinations
arrange themselves into the following 
multiplets of $A_5$:
\begin{equation}
\begin{aligned}
Y^{(4)}_\mathbf{1}    &\,=\, Y_1^2+2 Y_3 Y_4+2 Y_2 Y_5 \,\sim\, \mathbf{1}  \,,\\[2mm]
{Y^{(4)}_\mathbf{3}}  &\,=\, 
\left(
\begin{array}{c}
 -2 Y_1 Y_6+\sqrt{3}\, Y_5 Y_7+\sqrt{3}\, Y_2 Y_8 \\
 \sqrt{3}\, Y_2 Y_6+Y_1 Y_7-\sqrt{6}\, Y_3 Y_8 \\
 \sqrt{3}\, Y_5 Y_6-\sqrt{6}\, Y_4 Y_7+Y_1 Y_8 \\
\end{array}
\right)
\,\sim\, \mathbf{3}  \,,\\[2mm]
{Y^{(4)}_\mathbf{3'}}  &\,=\,
\left(
\begin{array}{c}
 \sqrt{3}\, Y_1 Y_6+Y_5 Y_7+Y_2 Y_8 \\
 Y_3 Y_6-\sqrt{2}\, Y_2 Y_7-\sqrt{2}\, Y_4 Y_8 \\
 Y_4 Y_6-\sqrt{2}\, Y_3 Y_7-\sqrt{2}\, Y_5 Y_8 \\
\end{array}
\right)
\,\sim\, \mathbf{3'} \,,\\[2mm]
{Y^{(4)}_\mathbf{4}}  &\,=\,
\left(
\begin{array}{c}
 2 Y_4^2+\sqrt{6}\, Y_1 Y_2-Y_3 Y_5 \\
 2 Y_2^2+\sqrt{6}\, Y_1 Y_3-Y_4 Y_5 \\
 2 Y_5^2-Y_2 Y_3+\sqrt{6}\, Y_1 Y_4 \\
 2 Y_3^2-Y_2 Y_4+\sqrt{6}\, Y_1 Y_5 \\
\end{array}
\right)
\,\sim\, \mathbf{4}\,,\\[2mm]
{Y^{(4)}_{\mathbf{5},1}}  &\,=\,
\left(
\begin{array}{c}
 \sqrt{2}\, Y_1^2+\sqrt{2}\, Y_3 Y_4-2 \sqrt{2}\, Y_2 Y_5 \\
 \sqrt{3}\, Y_4^2-2 \sqrt{2}\, Y_1 Y_2 \\
 \sqrt{2}\, Y_1 Y_3+2 \sqrt{3}\, Y_4 Y_5 \\
 2 \sqrt{3}\, Y_2 Y_3+\sqrt{2}\, Y_1 Y_4 \\
 \sqrt{3}\, Y_3^2-2 \sqrt{2}\, Y_1 Y_5 \\
\end{array}
\right)
\,\sim\, \mathbf{5}\,,\\[2mm]
{Y^{(4)}_{\mathbf{5},2}}  &\,=\,
\left(
\begin{array}{c}
 \sqrt{3}\, Y_5 Y_7-\sqrt{3}\, Y_2 Y_8 \\
 -Y_2 Y_6-\sqrt{3}\, Y_1 Y_7-\sqrt{2}\, Y_3 Y_8 \\
 -2 Y_3 Y_6-\sqrt{2}\, Y_2 Y_7 \\
 2 Y_4 Y_6+\sqrt{2}\, Y_5 Y_8 \\
 Y_5 Y_6+\sqrt{2}\, Y_4 Y_7+\sqrt{3}\, Y_1 Y_8 \\
\end{array}
\right)
\,\sim\, \mathbf{5}\,.
\label{eq:21}
\end{aligned}
\end{equation}
%

Finally, we show in Appendix~\ref{app:Dedekind} that relying on the Dedekind eta function for construction of the seed functions (the approach previously employed in the literature) is not enough to fully capture the results reported in eq.~\eqref{eq:seed}.

\section{Phenomenology}
\label{sec:pheno}
To illustrate the use of the constructed modular multiplets for model building, we consider a minimal example where the neutrino masses originate from the Weinberg operator.
We assume that the charged lepton mass matrix is diagonal, 
so it does not contribute to the mixing.
We will show later an explicit example with residual symmetry 
where this possibility is realised.
It should be mentioned that here we do not aim to construct 
a ``perfect'' minimal model and/or to  perform an exhaustive analysis of  
possible phenomenologically viable models based on the modular $A_5$ symmetry.
This will be done elsewhere. Our goal is only to present a
``proof of existence'' of such models by constructing 
a simple model which describes the neutrino data
and possibly makes testable predictions.

In the set-up outlined above, 
the only superpotential term relevant for the mixing is the Weinberg operator:
\begin{align}
W \supset
\frac{g}{\Lambda}\,\left( L \,H_u\, L\, H_u\, Y \right)_\1\,,
\end{align}
%
where $Y$ is a modular multiplet of weight $k_Y$.

We assume that the lepton $\text{SU}(2)_L$ doublets transform as an $A_5$ triplet ($\r_L \sim \3$ or $\tp$) of weight $-k_L$, while the Higgs multiplet $H_u$ is an $A_5$ singlet ($\r_u \sim \1$) of zero weight ($k_u = 0$).
After the breaking of the modular symmetry, we obtain:
\begin{align}
\frac{g}{\Lambda}\,\left( L \,H_u\, L\, H_u\, Y \right)_\1\, \rightarrow\,
c_{ij}\,(L_i \,H_u)\,(L_j \,H_u)\,,
\end{align}
%
which leads to the Lagrangian term
\begin{align}
\mathcal{L} \,\supset\,
-\frac{1}{2}\,\big(M_\nu\big)_{ij}\,\overline{\nu_{iR}^c}\,\nu_{jL}
+ \text{h.c.}
\,,
\end{align}
%
written in terms of four-spinors,
where
$M_\nu \equiv 2\, c\, v_u^2$,
with $\langle H_u\rangle = (0,v_u)^T$, 
and $\nu_{iR}^c \equiv (\nu_{iL})^c \equiv C \overline{\nu_{iL}}^T$, with $C$ being the charge conjugation matrix.

Given the above conditions, one needs to have $k_Y = 2 k_L$ to compensate the overall weight of the Weinberg operator term.
Since $k_Y$ is a non-negative integer, we can systematically explore the possible neutrino mass matrices going from $k_Y = 0$ to more and more positive integer $k_Y$. 
In the case of  $k_Y = 0$ there are no modular forms in the Weinberg operator and the only possible $A_5$ singlet is $ (LL)_\1 = L_1 L_1 + L_2 L_3 + L_3 L_2$ (cf.~Appendix~\ref{app:cbc}),
which leads to the following neutrino mass matrix:
\begin{align}
M_\nu = \frac{2 v_u^2 g}{\L}
\begin{pmatrix}
1 & 0 & 0\\0 & 0 & 1\\0 & 1 & 0 
\end{pmatrix}\,.
\end{align}
The case $k_Y = 0$ is then excluded, since it leads to degenerate neutrino masses, which are ruled out by the neutrino oscillation data \cite{PDG2018}. In the following subsections we consider the cases $k_Y = 2$ and $k_Y = 4$ (corresponding to $k_L = 1$ and $k_L = 2$, respectively).

\subsection{The case of  \texorpdfstring{$k_Y=2$}{kY=2}}
In this case, the available modular form multiplets are $Y_\3$, $Y_\tp$ and $Y_\5$.
Note that $L^2$ decomposes as $\3^{(\prime)} \ot \3^{(\prime)} = \1 \oplus \3^{(\prime)} \oplus \5$, but the $\3^{(\prime)}$ component vanishes due to antisymmetry (see Appendix~\ref{app:cbc}).
Therefore the only way to form a singlet is by combining the quintets, $\big(L^2\,Y_{\mathbf{5}}\big)_\mathbf{1}$.
If $\r_L \sim \3$, one obtains:
\begin{equation}
M_\nu^{\3}
=\frac{v_u^2 g}{\Lambda}
\begin{pmatrix}
2 Y_1         & -\sqrt{3} Y_5   & -\sqrt{3} Y_2   \\
-\sqrt{3} Y_5 & \sqrt{6} Y_4 & -Y_1            \\
-\sqrt{3} Y_2 & -Y_1            & \sqrt{6} Y_3    
\end{pmatrix}
\,,
\label{eq:k2Mnu3}
\end{equation}
while if instead $\r_L \sim \tp$, it follows that
\begin{equation}
M_\nu^{\tp}
=\frac{v_u^2 g}{\Lambda}
\begin{pmatrix}
2 Y_1         & -\sqrt{3} Y_4   & -\sqrt{3} Y_3   \\
-\sqrt{3} Y_4 & \sqrt{6} Y_2    & -Y_1            \\
-\sqrt{3} Y_3 & -Y_1            & \sqrt{6} Y_5    
\end{pmatrix}
\,.
\label{eq:k2Mnu3p}
\end{equation}
The difference between eq.~\eqref{eq:k2Mnu3} and eq.~\eqref{eq:k2Mnu3p} resides in the cyclic exchange of $Y_5$, $Y_4$, $Y_2$ and $Y_3$ (in this order).

In both cases, $\langle \tau \rangle$ determines neutrino masses up to an overall mass scale. Furthermore, given our assumption of a diagonal charged lepton mass matrix, 
after employing the permutation ordering the charged lepton masses,
$\langle \tau \rangle$ additionally determines the mixing parameters.
Through numerical search, we find that the agreement with data is optimised by choosing $\r_L = \tp$ and $\langle \tau \rangle = 0.48 + 0.873\, i$, giving rise to the following values of observables,%
\footnote{
Deriving the VEV of $\tau$ from a potential is 
out of the scope of our study.
Here we treat  $\langle \tau \rangle$ as a free parameter determined by fits to the data.  
} 
for a spectrum with normal ordering:
\begin{equation}
\begin{gathered}
r = 0.03056,\quad
\Delta m^2_{21} = 7.427 \cdot 10^{-5} \text{ eV}^2,\quad
\Delta m^2_{31} = 2.467 \cdot 10^{-3} \text{ eV}^2,\\
m_1 = 0.02036 \text{ eV},\quad
m_2 = 0.02211 \text{ eV},\quad
m_3 = 0.05368 \text{ eV},\quad
\textstyle\sum_i m_i = 0.09616 \text{ eV},\\
\sin^2 \theta_{12} = 0.3252,\quad
\sin^2 \theta_{13} = {\color{red} 0.1655},\quad
\sin^2 \theta_{23} = 0.4213,\\
\delta / \pi = 1.498,\quad
\alpha_{21} / \pi = 1.904,\quad
\alpha_{31} / \pi = 1.948,
\end{gathered}
\end{equation}
given an overall factor $v_u^2\, g / \Lambda \simeq 0.006339$ eV, and assuming the charged lepton sector induces a permutation of the first and third rows of the PMNS mixing matrix.
While one obtains a good agreement with data for the mass-squared differences (and hence for the ratio $r \equiv \Delta m^2_{21} / \Delta m^2_{31}$), as well as for the values of $\sin^2 \theta_{12}$ and of $\delta$, the value of $\sin^2 \theta_{23}$ is slightly outside its $3\sigma$ range and, more importantly, $\sin^2 \theta_{13}$ is many standard deviations away from its experimentally allowed range~\cite{Capozzi:2018ubv,Esteban:2018azc}. 
Nevertheless,
it is encouraging to find that the predictions for the mixing angles are in qualitative agreement with the observed pattern, namely, $\sin^2 \theta_{13} < \sin^2 \theta_{12} < \sin^2 \theta_{23}$.
Note that the indicated value of $\langle\tau\rangle$ is close to the ``right cusp'' $\tau_R =  1/2 + i\,\sqrt{3}/2$, which preserves a residual $\mathbb{Z}_3^{TS}$ symmetry (see, e.g.,~\cite{Novichkov:2018ovf}).

\subsection{The case of \texorpdfstring{$k_Y=4$}{kY=4}}
In this case, the available modular form multiplets are those given in eq.~\eqref{eq:21}.
Again, since $L^2$ decomposes as $\1 \oplus \5$, one can form singlets by using $Y^{(4)}_\mathbf{1}$, $Y^{(4)}_{\mathbf{5},1}$ or $Y^{(4)}_{\mathbf{5},2}$. All three contributions should enter $W$ with  independent complex coefficients.
If $\r_L \sim \3$, one obtains:
\begin{align}
M_\nu^{\mathbf{3}} &= 
\frac{2v_u^2 g_1}{\Lambda}\left[
(Y_1^2+2Y_3Y_4+2Y_2Y_5)\begin{pmatrix}
1&0&0\\0&0&1\\0&1&0
\end{pmatrix}\right.\nonumber\\[2mm]
&+
\frac{g_2}{g_1}\begin{pmatrix}
Y_1^2+Y_3Y_4-2Y_2Y_5  & -\frac{3}{2\sqrt{2}} Y_3^2 + \sqrt{3} Y_1 Y_5 & 
-\frac{3}{2\sqrt{2}} Y_4^2 + \sqrt{3} Y_1 Y_2 \\
\ast & 3Y_2 Y_3+\sqrt{\frac{3}{2}}Y_1 Y_4  &
Y_2 Y_5 - \frac{1}{2}(Y_1^2 + Y_3 Y_4) \\
\ast & \ast &  
3Y_4 Y_5+\sqrt{\frac{3}{2}}Y_1 Y_3  
\end{pmatrix}\\[2mm] \nonumber
&\left.
+
\frac{g_3}{g_1}
\begin{pmatrix}
Y_5 Y_7-Y_2 Y_8 & -\frac{1}{2} Y_5 Y_6 - {\frac{1}{\sqrt2}} Y_4 Y_7 -\frac{\sqrt3}{2} Y_1 Y_8 &
\frac{1}{2} Y_2 Y_6 + \frac{\sqrt3}{2} Y_1 Y_7 + \frac{1}{\sqrt2} Y_3 Y_8\\
\ast & \sqrt{2} Y_4 Y_6 + Y_5 Y_8 &     \frac{1}{2} (Y_2 Y_8 - Y_5 Y_7) \\
\ast & \ast & -\sqrt{2}Y_3 Y_6 - Y_2 Y_7
\end{pmatrix}
\right],
\end{align}
where through asterisks we (here and henceforth) omit some entries of symmetric matrices.
If instead $\r_L \sim \tp$, it follows that
\begin{align}
M_\nu^{\mathbf{3}'} &= 
\frac{2v_u^2 g_1}{\Lambda}\left[
(Y_1^2+2Y_3Y_4+2Y_2Y_5)\begin{pmatrix}
1&0&0\\0&0&1\\0&1&0
\end{pmatrix}\right. \nonumber \\[2mm]
& +
\frac{g_2}{g_1}\begin{pmatrix}
Y_1^2+Y_3Y_4-2Y_2Y_5  & 
-\frac{3}{\sqrt{2}}Y_2 Y_3-\frac{\sqrt{3}}{2}Y_1 Y_4 &
-\frac{\sqrt{3}}{2}Y_1 Y_3 - \frac{3}{\sqrt{2}} Y_4 Y_5\\
\ast &
\frac{3}{2} Y_4^2 - \sqrt{6} Y_1 Y_2 &
Y_2 Y_5 - \frac{1}{2}(Y_1^2 + Y_3 Y_4) \\
\ast & 
\ast &  
\frac{3}{2} Y_3^2 - \sqrt{6} Y_1 Y_5
\end{pmatrix}\\[2mm]  \nonumber
&\left.+
\frac{g_3}{g_1}
\begin{pmatrix}
Y_5 Y_7-Y_2 Y_8 &
- Y_4 Y_6 -{\frac{1}{\sqrt{2}}} Y_5 Y_8  &
Y_3 Y_6 + {\frac{1}{\sqrt{2}}} Y_2 Y_7 \\
\ast  &
-{\frac{1}{\sqrt{2}}} Y_2 Y_6 -\sqrt{\frac{3}{2}} Y_1 Y_7 - Y_3 Y_8&
\frac{1}{2} (Y_2 Y_8 - Y_5 Y_7) \\
\ast &
\ast &
{\frac{1}{\sqrt{2}}} Y_5 Y_6 +  Y_4 Y_7 + \sqrt{\frac{3}{2}} Y_1 Y_8
\end{pmatrix}
\right]\,.
\end{align}

\subsection{The case of residual symmetries}
Let us assume that the modular symmetry is broken to the residual 
$\mathbb{Z}_2^S = \left\{ I, S \right\}$ symmetry 
of the neutrino mass matrix.
This can be achieved by fixing $\langle \tau \rangle = i$ 
in the neutrino sector,
since this value of $\vev{\t}$ 
is invariant under the action of $S$, $\tau \to - 1/\tau$.

For the symmetric value of the VEV of the modulus 
$\langle \tau \rangle = i$,
the lowest weight modular forms $Y_i$ take the values:
\begin{equation}
\begin{gathered}
Y_2 = \frac{-1-\sqrt{7-4\varphi}}{\sqrt{6}}\,Y_1\,,~~~
Y_3 = \frac{-1-\sqrt{18-11\varphi}}{\sqrt{6}}\,Y_1\,,\\
Y_4 = \frac{-1+\sqrt{18-11\varphi}}{\sqrt{6}}\,Y_1\,,~~~
Y_5 = \frac{-1+\sqrt{7-4\varphi}}{\sqrt{6}}\,Y_1\,,\\
Y_6 = \sqrt{\frac{58-31\varphi}{15}}\,Y_1\,,~~~
Y_7 = \frac{-9+8\varphi+\sqrt{27-4\varphi}}{\sqrt{30}}\, Y_1\,,~~~
Y_8 = \frac{9-8\varphi+\sqrt{27-4\varphi}}{\sqrt{30}}\, Y_1\,,\\
Y_9 = -\sqrt{\frac{3+4\varphi}{15}}\,Y_1\,,~~~
Y_{10} = \frac{7-4\varphi+\sqrt{2+\varphi}}{\sqrt{30}}\, Y_1\,,~~~
Y_{11} = \frac{-7+4\varphi+\sqrt{2+\varphi}}{\sqrt{30}}\, Y_1\,,
\end{gathered}
\end{equation}
with $Y_1 (\langle \tau \rangle = i) \simeq 2.594\,i$.

It follows that in both cases $\rho_L \sim \3,\tp$ for $k_Y = 4$, 
aside from the permutation which orders charged lepton masses, 5 real parameters~---~namely, $2v_u^2\, g_1 / \L$, $\re (g_2/g_1)$, $\im (g_2/g_1)$, $\re (g_3/g_1)$ and $\im (g_3/g_1)$~---~determine the neutrino masses and mixing.
Through numerical search, we find 
for $\rho_L \sim \tp$
a point given by:
\begin{equation}
2v_u^2\, g_1 / \Lambda \simeq 0.005104~\text{eV},\quad
g_2 / g_1 = -0.2205 - 0.1576\, i,\quad
g_3 / g_1 = 0.0246- 0.0421\, i,
\label{eq:benchmark}
\end{equation}
which is consistent with the neutrino oscillation data at $1.7\sigma$ level,%
\footnote{The current upper limit on the sum of neutrino masses reported in 2018 by the Planck collaboration depends on the data set used as input and reads~\cite{Aghanim:2018eyx}: $\sum_i m_i < 0.120 - 0.160$ eV at 95\% C.L.}
for a spectrum with normal ordering:
\begin{equation}
\begin{gathered}
r = 0.03,\quad
\Delta m_{21}^2 = 7.399 \cdot 10^{-5} \text{ eV}^2,\quad
\Delta m_{31}^2 = 2.489 \cdot 10^{-3} \text{ eV}^2,\\
m_1 = 0.0416 \text{ eV},\quad
m_2 = 0.04248 \text{ eV},\quad
m_3 = 0.06496 \text{ eV},\\
\textstyle\sum_i m_i = 0.149 \text{ eV},\quad
| \langle m \rangle | = 0.04174 \text{ eV},\\
\sin^2 \theta_{12} = 0.2824,\quad
\sin^2 \theta_{13} = 0.02136,\quad
\sin^2 \theta_{23} = 0.5504,\\
\delta / \pi = 1.315,\quad
\alpha_{21} / \pi = 1.978,\quad
\alpha_{31} / \pi = 0.9312.
\end{gathered}
\label{eq:benchmarkdata}
\end{equation}
%
The indicated value of the effective Majorana mass $|\langle m \rangle|$ which controls the rate of neutrinoless double 
beta decay may be probed in future experiments aiming to test values down to the $|\langle m \rangle| \sim 10^{-2}$ eV frontier.

\begin{figure}
    \centering
    \includegraphics[width=\textwidth]{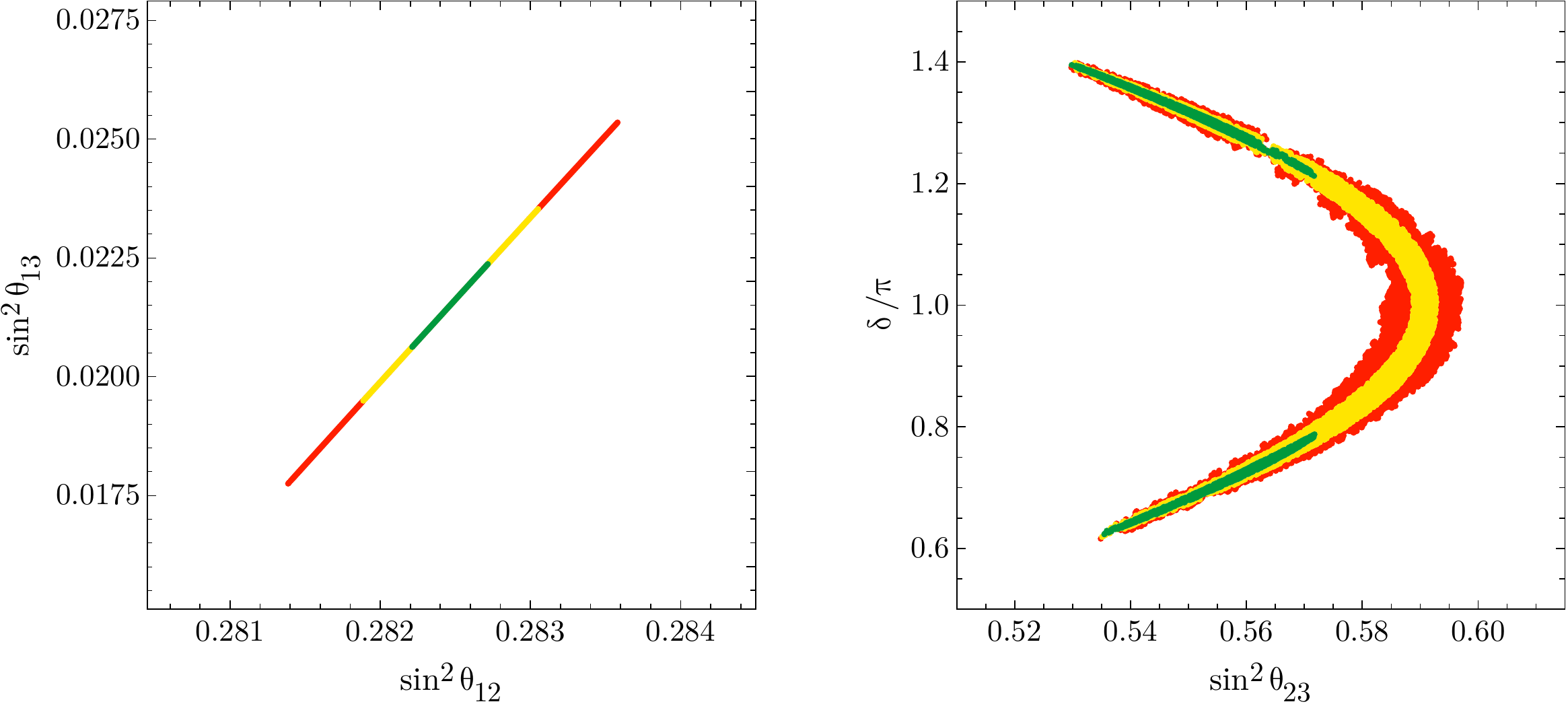}
    \caption{Correlations between $\sin^2 \theta_{12}$ and $\sin^2 \theta_{13}$ (left) and between $\sin^2 \theta_{23}$ and $\delta$ (right) in the model with $k_Y = 4$ and $\langle \tau \rangle = i$, in the vicinity of the viable point of eq.~\eqref{eq:benchmark}.
    The green, yellow and red regions correspond to $2\sigma$, $3\sigma$ and $5\sigma$ confidence levels, respectively.}
    \label{fig:corr}
\end{figure}

In the vicinity of the point described by eq.~\eqref{eq:benchmark}, keeping $\langle \tau \rangle = i$, we find strong correlations between $\sin^2 \theta_{12}$ and $\sin^2 \theta_{13}$, and between $\sin^2 \theta_{23}$ and $\delta$. These correlations are shown in Fig.~\ref{fig:corr}.

One possible way to force the charged lepton mass matrix to be diagonal in this set-up is to assume that it originates from a different modulus $\tau^l$ which develops a VEV $\langle \tau^l \rangle = i\, \infty$ breaking the modular symmetry to the residual $\mathbb{Z}$ symmetry generated by $T$, $\t \to \t + 1$.%
\footnote{It is not clear how the two moduli $\tau$
and $\tau^l$ can be  forced to couple either
to neutrinos or charged leptons only, so this possibility is considered
on purely phenomenological grounds.}
The corresponding residual symmetry of the charged lepton mass matrix is $\mathbb{Z}_5$ generated by the $T$ generator of $\G_5$, which is diagonal for $\rho_L \sim \3^{(\prime)}$.
One can show that in the case $\rho_{E^c} = \rho_L$ the charged lepton Yukawa interaction terms $(E^c L Y)_\1 H_d$ with the multiplets of weight 4 modular forms lead to the following mass matrix
(written in the left-right convention):%
\footnote{Actually, this is the most general form of the mass matrix for any weight higher than 2, since modular form singlets, triplets and quintets are always present at such weights, and their values at $\tau = i\,\infty$ are such that only their first components can be non-zero, cf.~eq.~\eqref{eq:Yinf}.
At weight 2, however, the first term in eq.~\eqref{eq:Me} is missing as there is no modular form singlet of weight 2, and it is impossible to recover the charged lepton mass hierarchy in this case.}
\begin{equation}
M_e = v_d\, \alpha_1 \left[
\begin{pmatrix} 1 & 0 & 0 \\ 0 & 0 & 1 \\ 0 & 1 & 0 \end{pmatrix}
+ \frac{\alpha_2}{\alpha_1}
\begin{pmatrix} 0 & 0 & 0 \\ 0 & 0 & -1 \\ 0 & 1 & 0 \end{pmatrix}
+ \frac{\alpha_3}{\alpha_1}
\begin{pmatrix} 2 & 0 & 0 \\ 0 & 0 & -1 \\ 0 & -1 & 0 \end{pmatrix}
\right]
\,,
\label{eq:Me}
\end{equation}
where the three matrix terms correspond to contributions from $Y_\1^{(4)}$, $Y_{\3^{(\prime)}}^{(4)}$ and $Y_{\5,1}^{(4)}$ respectively, and $\vev{H_d} = (v_d, 0)^T$.
The relevant product is diagonal:
\be
M_e M_e^\dagger = v_d^2\, \alpha_1^2 \diag \left(
\left| 1+2\,\frac{\alpha_3}{\alpha_1} \right|^2,\,
\left| 1-\frac{\alpha_2}{\alpha_1}-\frac{\alpha_3}{\alpha_1} \right|^2,\,
\left| 1+\frac{\alpha_2}{\alpha_1}-\frac{\alpha_3}{\alpha_1} \right|^2
\right),
\ee
where $\alpha_2 / \alpha_1$ and $\alpha_3 / \alpha_1$ are complex parameters, and $v_d\, \alpha_1$ is the overall mass scale factor.
To reproduce the charged lepton masses with these parameters one can choose, e.g., $v_d\, \alpha_1 \simeq 660$~MeV, $\alpha_2 / \alpha_1 = 1.34$ and $1 + 2 \alpha_3 / \alpha_1 = -7.7 \times 10^{-4}$.
It is interesting to note that all three $\alpha_i$
are of the same order, $\alpha_1 \sim\alpha_2 \sim\alpha_3$.

It is also possible to obtain the same matrix without an additional modulus $\tau^l$.
Instead, suppose that the combination $E^c L$ has positive modular weight, so that it cannot couple to modular forms.
Let us assume that the charged lepton mass matrix originates from 
Yukawa couplings to three flavons, an $A_5$ singlet $\varphi_{\1}$, an $A_5$ triplet $\varphi_{\3^{(\prime)}}$ and an $A_5$ quintet $\varphi_\5$ of a negative weight opposite to that of $E^c L$, each of which develops a VEV breaking $A_5$ to $\mathbb{Z}_5^T$:
\begin{equation}
\langle \varphi_{\1} \rangle = v_1,\quad
\langle \varphi_{\3^{(\prime)}} \rangle = (v_2, 0, 0)\quad\text{and}\quad
\langle \varphi_\5 \rangle = (v_3, 0, 0, 0, 0).
\label{eq:flavonVEV}
\end{equation}
In this case the three terms $(E^c L \varphi_{\1})_\1 H_d$, $(E^c L \varphi_{\3^{(\prime)}})_\1 H_d$ and $(E^c L \varphi_\5)_\1 H_d$ lead to the same mass matrix as in eq.~\eqref{eq:Me}.
This is related to the fact that modular form multiplets of weight 4 take the following values at the symmetric point $\langle \tau \rangle = i\, \infty$:
\begin{equation}
Y_\3^{(4)} = \frac{4\pi^2}{\sqrt{15}} (1, 0, 0),\quad
Y_\tp^{(4)} = -\frac{2\pi^2}{\sqrt{5}} (1, 0, 0),\quad
Y_{\5,1}^{(4)} = -\frac{2\sqrt{2} \pi^2}{3} (1, 0, 0, 0, 0).
\label{eq:Yinf}
\end{equation}
Hence, they can be thought of as flavon multiplets developing the corresponding VEVs.

The above-described construction of the charged lepton mass matrix has been considered earlier in Ref.~\cite{Ding:2011cm}, where the authors provide also an explicit form of the flavon potential which leads to the VEVs given in eq.~\eqref{eq:flavonVEV}.
An auxiliary $\mathbb{Z}_5$ symmetry is used in Ref.~\cite{Ding:2011cm} to forbid certain terms in the superpotential.
This is not needed in our case, since we do not introduce flavons which couple to the neutrino sector, and the $(E^c L)_{\1} H_d$ term is forbidden in our construction by a choice of a non-zero modular weight.
As a final remark, the parameters $\alpha_i$ of the charged lepton mass matrix~\eqref{eq:Me} have to be fine-tuned in order to reproduce the observed mass hierarchies.
To overcome this defect, the authors of Ref.~\cite{Ding:2011cm} propose also an alternative model, which can be adopted directly in our case.

\section{Summary and Conclusions}
\label{sec:conclusions}
 In the framework of the modular invariance approach 
to lepton flavour proposed in Ref.~\cite{Feruglio:2017spp}, 
we have considered a class of theories
in which couplings and matter superfields 
transform in irreps of the finite modular group 
$\Gamma_5 \simeq A_5$. 
The building blocks needed to construct such
theories are modular forms of weight 2 and level 5. 
We have explicitly constructed the 
11 generating modular forms of weight 2, 
using the Jacobi theta function 
and its properties, which lead to closure 
of the set of 12 seed functions (see eq.~\eqref{eq:seed}) 
under the action of $\G_5$, as shown in Fig.~\ref{fig:graph}.
Further, we have demonstrated how these 
11 modular forms arrange themselves 
into multiplets of $A_5$. 
Namely, we have found two triplets transforming in the 
irreps $\3$ and $\tp$ of $A_5$, 
and a quintet transforming in the irrep $\5$ of $A_5$. 
They are given in eqs.~\eqref{eq:5}\,--\,\eqref{eq:3p}, and their explicit $q$-expansions are listed in Appendix~\ref{app:qexp}. 
From these triplets and quintet
we have constructed multiplets of modular forms 
of weight 4 (see eq.~\eqref{eq:21} and Appendix~\ref{app:constraints}).

While thorough analysis of modular-invariant 
theories with the $\G_5$ symmetry is left 
for future work, we have presented two examples 
of application of the obtained results 
to neutrino masses and mixing. 
In both of them, we have assumed that 
neutrino masses are generated via
the Weinberg operator, and 
considered the charged lepton mass matrix to be diagonal.
The first model involving the quintet of 
weight 2 modular forms leads to 
the neutrino mass matrix containing 
three real parameters~---~complex VEV of the modulus 
$\langle \t \rangle$ and a real overall scale. 
We have found that the value of $\langle\t\rangle$ 
lying very close to the ``right cusp" 
$\t_R = 1/2 + i\,\sqrt3/2$ 
($\t_R$ preserves a residual $\mathbb{Z}_3^{TS}$ symmetry)
leads to a good agreement with neutrino oscillation data 
except for $\sin^2\th_{13}$, 
which falls many standard deviations away 
from its experimentally allowed region. 
The second model contains an $A_5$ singlet and two 
$A_5$ quintets of modular forms of weight~4. 
The neutrino mass matrix in this case depends 
on five real parameters 
(three real constants and two phases)
apart from $\vev{\t}$.
Assuming that $\vev{\t} = i$~---~%
a self-dual point which preserves 
a residual $\mathbb{Z}_2^S$ symmetry~---~%
we have obtained a viable benchmark point 
compatible with the data at $1.7\sigma$ confidence level. 
In this case the neutrino mass matrix depends on three 
real parameters and two phases.
Varying these free parameters
we found strong correlations 
between the values of $\sin^2\th_{12}$ and $\sin^2\th_{13}$, 
and the values of $\sin^2\th_{23}$ and the Dirac CPV phase $\delta$ 
(Fig.~\ref{fig:corr}). 

 In conclusion, the results obtained in the 
present study can be used to 
build in a systematic way modular-invariant 
flavour models with the $\G_5 \simeq A_5$ symmetry.  
In this regard, this article is expected to serve 
as a useful handbook for future studies.

\section*{Acknowledgements}
A.V.T.~would like to thank F.~Feruglio 
for useful discussions on 
problems related to this work. 
A.V.T.~expresses his gratitude to SISSA and 
the University of Padua, where part of this 
work was carried out, 
for their hospitality and support. 
This project has received funding from the European Union's Horizon 2020 research and innovation programme under the Marie 
Skłodowska-Curie grant agreements No 674896 
(ITN Elusives) and No 690575 (RISE InvisiblesPlus).
This work was supported in part 
by the INFN program on Theoretical Astroparticle Physics (P.P.N. and S.T.P.)
and by the  World Premier International Research Center
Initiative (WPI Initiative, MEXT), Japan (S.T.P.).
The work of J.T.P. was partially supported by Fundação para a Ciência e a Tecnologia (FCT,  Portugal) through the project CFTP-FCT Unit 777 (UID/FIS/00777/2013) and PTDC/FIS-PAR/29436/2017 which are partially funded through POCTI (FEDER), COMPETE, QREN and EU.

\appendix
\section{\texorpdfstring{$q$}{q}-Expansions}
\subsection{Miller-like basis for the space of lowest weight forms}
\label{app:sage}
One can obtain $q$-expansions for a basis of the space of lowest weight modular forms for $\Gamma_5$ from the SageMath algebra system~\cite{SageMath:2018}.
To obtain the expansions up to (and including) $\mathcal{O}({q^{10})}$ terms, we take as input the code:
\lstset{frame=tb,
  aboveskip=3mm,
  belowskip=3mm,
  showstringspaces=false,
  columns=flexible,
  basicstyle={\small\ttfamily},
  numbers=none,
  breaklines=true,
  breakatwhitespace=true,
  tabsize=3
}
\begin{lstlisting}
N=5
k=2
space = ModularForms(GammaH(N^2, [N + 1]), k)
[form.q_expansion(51) for form in space.basis()]
\end{lstlisting}
We then take $q \rightarrow q^{1/5}$ in the produced output,
and obtain the desired basis:
\begin{equation}
\begin{aligned}
b_1 &= 1 + 60 q^3 - 120 q^4 + 240 q^5 - 300 q^6 + 300 q^7 - 180 q^9 + 
  240 q^{10}+\ldots\,,\\
b_2 &=  q^{1/5} + 12 q^{11/5} + 7 q^{16/5} + 8 q^{21/5} + 6 q^{26/5} + 
  32 q^{31/5} + 7 q^{36/5} + 42 q^{41/5} + 12 q^{46/5}+\ldots\,, \\
b_3 &=  q^{2/5} + 12 q^{12/5} - 2 q^{17/5} + 12 q^{22/5} + 8 q^{27/5} + 
  21 q^{32/5} - 6 q^{37/5} + 48 q^{42/5} - 8 q^{47/5}+\ldots\,, \\
b_4 &=  q^{3/5} + 11 q^{13/5} - 9 q^{18/5} + 21 q^{23/5} - q^{28/5} + 
  12 q^{33/5} + 41 q^{43/5} - 29 q^{48/5}+\ldots\,, \\
b_5 &=  q^{4/5} + 9 q^{14/5} - 12 q^{19/5} + 29 q^{24/5} - 18 q^{29/5} + 
  17 q^{34/5} + 8 q^{39/5} + 12 q^{44/5} - 16 q^{49/5}+\ldots\,, \\
b_6 &=  q + 6 q^3 - 9 q^4 + 27 q^5 - 28 q^6 + 30 q^7 - 11 q^9 + 26 q^{10}+\ldots\,,\\ 
b_7 &=  q^{6/5} + 2 q^{16/5} + 2 q^{21/5} + 3 q^{26/5} + 7 q^{36/5} + 
  5 q^{46/5}+\ldots\,, \\
b_8 &=  q^{7/5} - q^{12/5} + 3 q^{17/5} + 2 q^{27/5} + 7 q^{37/5} - 
  6 q^{42/5} + 9 q^{47/5}+\ldots\,,\\ 
b_9 &=  q^{8/5} - 2 q^{13/5} + 5 q^{18/5} - 4 q^{23/5} + 4 q^{28/5} + 
  4 q^{38/5} - 8 q^{43/5} + 16 q^{48/5}+\ldots\,,\\ 
b_{10} &=  q^{9/5} - 3 q^{14/5} + 8 q^{19/5} - 11 q^{24/5} + 12 q^{29/5} - 
  5 q^{34/5} + 13 q^{49/5}+\ldots\,, \\
b_{11} &=  q^2 - 4 q^3 + 12 q^4 - 22 q^5 + 30 q^6 - 24 q^7 + 5 q^8 + 18 q^9 - 
  21 q^{10}+\ldots\,,
\end{aligned}
\end{equation}
with $q = e^{2\pi i\,\tau}$ and where fractional powers $q^{n/5}$ should be read as $q^{n/5} = e^{2 n \pi i\,\tau/5}$.

\subsection{Expansions for the lowest weight \texorpdfstring{$A_5$}{A5} multiplets}
\label{app:qexp}
The elements of the quintet $\mathbf{5}$ of $A_5$,
given in eq.~\eqref{eq:5}, admit the $q$-expansions:
\begin{equation}
\begin{aligned}
-\frac{i}{\pi}\sqrt{\frac{3}{2}}\, Y_1(\tau) &=
1 + 6 q + 18 q^2 + 24 q^3 + 42 q^4 + 6 q^5
+\ldots
= b_1 + 6 b_6 + 18 b_{11}
\,,\\[2mm]
\frac{i}{2\pi}\, Y_2(\tau) &=
q^{1/5}+12 q^{6/5}+12 q^{11/5}+31 q^{16/5}+32 q^{21/5}
+\ldots
= b_2 + 12 b_7
\,,\\[2mm]
\frac{i}{2\pi}\, Y_3(\tau) &=
3 q^{2/5}+8 q^{7/5}+28 q^{12/5}+18 q^{17/5}+36 q^{22/5}
+\ldots
= 3 b_3 + 8 b_8
\,,\\[2mm]
\frac{i}{2\pi}\, Y_4(\tau) &=
4 q^{3/5}+15 q^{8/5}+14 q^{13/5}+39 q^{18/5}+24 q^{23/5}
+\ldots
= 4 b_4 + 15 b_9
\,,\\[2mm]
\frac{i}{2\pi}\, Y_5(\tau) &=
7 q^{4/5}+13 q^{9/5}+24 q^{14/5}+20 q^{19/5}+60 q^{24/5}
+\ldots
= 7 b_5 + 13 b_{10}
\,,
\end{aligned}
\end{equation}
where, as before and in what follows, $q = e^{2\pi i\,\tau}$.

The elements of the triplet $\mathbf{3}$, 
given in eq.~\eqref{eq:3}, admit instead the expansions:
\begin{equation}
\begin{aligned}
\frac{i}{\pi}\sqrt{\frac{5}{2}}\, Y_6(\tau) &=
-1 + 30 q + 20 q^2 + 40 q^3 + 90 q^4 + 130 q^5
+\ldots
= -b_1 + 30 b_6 + 20 b_{11}
\,,\\[2mm]
-\frac{i}{2\sqrt{5}\pi}\, Y_7(\tau) &=
q^{1/5}+2 q^{6/5}+12 q^{11/5}+11 q^{16/5}+12 q^{21/5}
+\ldots
= b_2 + 2 b_7
\,,\\[2mm]
-\frac{i}{2\sqrt{5}\pi}\, Y_8(\tau) &=
3 q^{4/5}+7 q^{9/5}+6 q^{14/5}+20 q^{19/5}+10 q^{24/5}
+\ldots
= 3b_5 + 7 b_{10}
\,.
\end{aligned}
\end{equation}

Finally, the elements of the triplet $\mathbf{3'}$,
given in eq.~\eqref{eq:3p}, read:
\begin{equation}
\begin{aligned}
\frac{i}{\pi}\sqrt{\frac{5}{2}}\, Y_9(\tau) &=
1 + 20 q + 30 q^2 + 60 q^3 + 60 q^4 + 120 q^5+\ldots
= b_1 + 20 b_6 + 30 b_{11}
\,,\\[2mm]
-\frac{i}{2\sqrt{5}\pi}\, Y_{10}(\tau) &=
q^{2/5}+6 q^{7/5}+6 q^{12/5}+16 q^{17/5}+12 q^{22/5}+\ldots
= b_3 + 6 b_8
\,,\\[2mm]
-\frac{i}{2\sqrt{5}\pi}\, Y_{11}(\tau) &=
2 q^{3/5}+5 q^{8/5}+12 q^{13/5}+7q^{18/5}+22 q^{23/5}+\ldots
= 2b_4 + 5 b_9
\,.
\end{aligned}
\end{equation}

\section{\texorpdfstring{$A_5$}{A5} Group Theory}
\label{app:A5}
\subsection{Basis}
\label{app:basis}
$A_5$ is the group of even permutations of five objects. It contains $5!/2 = 60$ elements and admits five irreducible representations, namely $\mathbf{1}$,
$\mathbf{3}$, $\mathbf{3'}$, $\mathbf{4}$ and $\mathbf{5}$ (see, e.g.,~\cite{Ishimori:2010au}). It can be generated by two elements $S$ and $T$ satisfying
\begin{align}
S^2 \,=\, (ST)^3 \,=\, T^5 \,=\, I\,.
\label{eq:A5presentation}
\end{align}

We will employ the group theoretical results of Ref.~\cite{Ding:2011cm},
using in particular the following explicit basis
for the $A_5$ generators in different irreps:
\begin{align}
\mathbf{1}:&\quad \rho(S)= 1 ,\quad \rho(T)= 1\,,\\
\mathbf{3}:&\quad \rho(S)=
\frac{1}{\sqrt{5}}
\begin{pmatrix}
 1 & -\sqrt{2} & -\sqrt{2} \\
 -\sqrt{2} & -\varphi & 1/{\varphi} \\
 -\sqrt{2} & 1/{\varphi} & -\varphi
\end{pmatrix} 
,\quad \rho(T)=
\begin{pmatrix}
1 & 0 & 0 \\
0 & \zeta & 0 \\
0 & 0 & \zeta^4   
\end{pmatrix}  \,,\\
\mathbf{3'}:&\quad \rho(S)= 
\frac{1}{\sqrt{5}}
\begin{pmatrix}
 -1 & \sqrt{2} & \sqrt{2} \\
 \sqrt{2} & -1/\varphi & \varphi \\
 \sqrt{2} & \varphi & -1/\varphi
\end{pmatrix} 
,\quad \rho(T)=
\begin{pmatrix}
1 & 0 & 0 \\
0 & \zeta^2 & 0 \\
0 & 0 & \zeta^3   \end{pmatrix}\,,\\
\mathbf{4}:&\quad \rho(S)= 
\frac{1}{\sqrt{5}}
\begin{pmatrix}
 1 & 1/\varphi & \varphi & -1 \\
 1/\varphi & -1 & 1 & \varphi \\
 \varphi & 1 & -1 & 1/\varphi \\
 -1 & \varphi & 1/\varphi & 1 
\end{pmatrix} 
,\quad \rho(T)=
\begin{pmatrix}
\zeta & 0 & 0 & 0\\
0 & \zeta^2 & 0 &0 \\
0 & 0 & \zeta^3 & 0 \\
0 & 0 & 0 & \zeta^4 
\end{pmatrix}\,,\\
\mathbf{5}:&\quad \rho(S)= 
\frac{1}{5}
\begin{pmatrix}
-1 & \sqrt{6} & \sqrt{6} & \sqrt{6} & \sqrt{6} \\
 \sqrt{6} & 1/{\varphi^2} & -2 \varphi  & {2}/{\varphi } & \varphi ^2 \\
 \sqrt{6} & -2 \varphi  & \varphi ^2 & {1}/{\varphi ^2} & {2}/{\varphi } \\
 \sqrt{6} & {2}/{\varphi } & 1/{\varphi ^2} & \varphi ^2 & -2 \varphi  \\
 \sqrt{6} & \varphi ^2 & {2}/{\varphi } & -2 \varphi  & {1}/{\varphi ^2}
\end{pmatrix} 
,\quad \rho(T)=
\begin{pmatrix}
1 & 0 & 0 & 0 & 0\\
0 & \zeta & 0 & 0 & 0\\
0 & 0 & \zeta^2 & 0 &0 \\
0 & 0 & 0 & \zeta^3 & 0 \\
0 & 0 & 0 & 0 & \zeta^4 
\end{pmatrix}\,, 
\label{eq:irrep5}
\end{align}
where $\zeta = e^{2\pi i/5}$ and $\varphi = (1+\sqrt{5})/2$.

\subsection{Clebsch-Gordan coefficients}
\label{app:cbc}
For completeness, we reproduce here the nontrivial Clebsch-Gordan coefficients of Ref.~\cite{Ding:2011cm}, given in the above basis. Entries of each multiplet entering the tensor product are denoted by $\alpha_i$ and $\beta_i$.
\begin{align}
\begin{array}{@{}c@{{}\,\otimes\,{}}c@{{}\,\,=\,\,{}}ll@{}}
\mathbf{3}&\mathbf{3}&\mathbf{1}\,\oplus\, \mathbf{3}\,\oplus\, \mathbf{5} &
\left\{\begin{array}{@{}l@{\quad\sim\quad}l@{}}
\quad \mathbf{1}  & \alpha_1\beta_1+\alpha_2\beta_3+\alpha_3\beta_2\\[2mm]
\quad \mathbf{3}  & \begin{pmatrix} \alpha_2\,\beta_3-\alpha_3\,\beta_2\\
                                    \alpha_1\,\beta_2-\alpha_2\,\beta_1\\
                                    \alpha_3\,\beta_1-\alpha_1\,\beta_3
                    \end{pmatrix}\\[6mm]
\quad \mathbf{5}  & \begin{pmatrix}  2 \alpha_1 \beta_1-\alpha_2 \beta_3-\alpha_3 \beta_2 \\
                                     -\sqrt{3}\, \alpha_1 \beta_2-\sqrt{3}\, \alpha_2 \beta_1 \\
                                     \sqrt{6}\, \alpha_2 \beta_2 \\
                                     \sqrt{6}\, \alpha_3 \beta_3 \\
                                     -\sqrt{3}\, \alpha_1 \beta_3-\sqrt{3}\, \alpha_3 \beta_1
                    \end{pmatrix}
\end{array}\right.
\end{array}
\end{align}
%
%
%
\begin{align}
\begin{array}{@{}c@{{}\,\otimes\,{}}c@{{}\,\,=\,\,{}}ll@{}}
\mathbf{3}&\mathbf{3'}&\mathbf{4}\,\oplus\,\mathbf{5}&
\left\{\begin{array}{@{}l@{\quad\sim\quad}l@{}}
\quad \mathbf{4}   & \begin{pmatrix}  \sqrt{2}\, \alpha_2 \beta_1+\alpha_3 \beta_2 \\
                                     -\sqrt{2}\, \alpha_1 \beta_2-\alpha_3 \beta_3 \\
                                     -\sqrt{2}\, \alpha_1 \beta_3-\alpha_2 \beta_2 \\
                                      \sqrt{2}\, \alpha_3 \beta_1+\alpha_2 \beta_3
                     \end{pmatrix}\\[9mm]
\quad \mathbf{5}  & \begin{pmatrix} \sqrt{3}\, \alpha_1 \beta_1 \\
                                    \alpha_2 \beta_1-\sqrt{2}\, \alpha_3 \beta_2 \\
                                    \alpha_1 \beta_2-\sqrt{2}\, \alpha_3 \beta_3 \\
                                    \alpha_1 \beta_3-\sqrt{2}\, \alpha_2 \beta_2 \\
                                    \alpha_3 \beta_1-\sqrt{2}\, \alpha_2 \beta_3
                     \end{pmatrix}
\end{array}\right.
\\[23mm]
\mathbf{3}&\mathbf{4}&\mathbf{3'}\,\oplus\,\mathbf{4}\,\oplus\,\mathbf{5}&
\left\{\begin{array}{@{}l@{\quad\sim\quad}l@{}}
\quad \mathbf{3'}   & \begin{pmatrix}  -\sqrt{2}\, \alpha_2 \beta_4-\sqrt{2}\, \alpha_3 \beta_1 \\
                                        \sqrt{2}\, \alpha_1 \beta_2-\alpha_2 \beta_1+\alpha_3 \beta_3 \\
                                        \sqrt{2}\,\alpha_1 \beta_3+\alpha_2 \beta_2-\alpha_3 \beta_4
                     \end{pmatrix}\\[6mm]
\quad \mathbf{4}   & \begin{pmatrix}   \alpha_1 \beta_1-\sqrt{2}\,\alpha_3 \beta_2 \\
                                      -\alpha_1 \beta_2-\sqrt{2}\,\alpha_2 \beta_1 \\
                                       \alpha_1 \beta_3+\sqrt{2}\,\alpha_3 \beta_4 \\
                                      -\alpha_1 \beta_4+\sqrt{2}\,\alpha_2 \beta_3
                     \end{pmatrix}\\[9mm]
\quad \mathbf{5}   & \begin{pmatrix}  \sqrt{6}\,\alpha_2\beta_4-\sqrt{6}\, \alpha_3 \beta_1 \\
                                       2\sqrt{2}\,\alpha_1 \beta_1+2 \alpha_3 \beta_2\\
                                       -\sqrt{2}\,\alpha_1 \beta_2+\alpha_2 \beta_1+3\alpha_3 \beta_3 \\
                                      \sqrt{2}\, \alpha_1 \beta_3-3\alpha_2 \beta_2-\alpha_3 \beta_4\\
                                       -2\sqrt{2}\, \alpha_1 \beta_4-2 \alpha_2 \beta_3
                     \end{pmatrix}
\end{array}\right.
\\[30mm]
\mathbf{3}&\mathbf{5}&\mathbf{3}\,\oplus\,\mathbf{3'}\,\oplus\,\mathbf{4}\,\oplus\,\mathbf{5}&
\left\{\begin{array}{@{}l@{\quad\sim\quad}l@{}}
\quad \mathbf{3}   & \begin{pmatrix}
                          -2 \alpha_1 \beta_1+\sqrt{3}\,\alpha_2 \beta_5+\sqrt{3}\,\alpha_3 \beta_2 \\
                          \sqrt{3}\,\alpha_1 \beta_2+\alpha_2 \beta_1-\sqrt{6}\,\alpha_3 \beta_3 \\
                          \sqrt{3}\,\alpha_1 \beta_5-\sqrt{6}\,\alpha_2 \beta_4+\alpha_3 \beta_1
                     \end{pmatrix}\\[6mm]
\quad \mathbf{3'}   & \begin{pmatrix} 
                          \sqrt{3}\,\alpha_1 \beta_1+\alpha_2 \beta_5+\alpha_3 \beta_2 \\
                          \alpha_1 \beta_3-\sqrt{2}\,\alpha_2 \beta_2-\sqrt{2}\,\alpha_3 \beta_4 \\
                          \alpha_1 \beta_4-\sqrt{2}\,\alpha_2 \beta_3-\sqrt{2}\,\alpha_3 \beta_5
                     \end{pmatrix}\\[6mm]
\quad \mathbf{4}   & \begin{pmatrix}
                          2\sqrt{2}\,\alpha_1 \beta_2-\sqrt{6}\, \alpha_2 \beta_1+\alpha_3 \beta_3 \\
                          -\sqrt{2}\,\alpha_1 \beta_3+2\alpha_2 \beta_2-3 \alpha_3 \beta_4 \\
                           \sqrt{2}\,\alpha_1 \beta_4+3\alpha_2 \beta_3-2\alpha_3 \beta_5 \\
                          -2\sqrt{2}\,\alpha_1 \beta_5-\alpha_2 \beta_4+\sqrt{6}\,\alpha_3 \beta_1
                     \end{pmatrix}\\[9mm]
\quad \mathbf{5}   & \begin{pmatrix}
                          \sqrt{3}\, \alpha_2 \beta_5-\sqrt{3}\, \alpha_3 \beta_2 \\
                          -\alpha_1 \beta_2-\sqrt{3}\,\alpha_2 \beta_1-\sqrt{2}\,\alpha_3 \beta_3 \\
                          -2 \alpha_1 \beta_3-\sqrt{2}\,\alpha_2 \beta_2 \\
                          2\alpha_1 \beta_4+\sqrt{2}\,\alpha_3 \beta_5 \\
                          \alpha_1 \beta_5+\sqrt{2}\,\alpha_2 \beta_4+ \sqrt{3}\,\alpha_3 \beta_1
                     \end{pmatrix}
\end{array}\right.
\end{array}
\end{align}

\begin{align}
\begin{array}{@{}c@{{}\,\otimes\,{}}c@{{}\,\,=\,\,{}}ll@{}}
\mathbf{3'}&\mathbf{3'}&\mathbf{1}\,\oplus\,\mathbf{3'}\,\oplus\,\mathbf{5}&
\left\{\begin{array}{@{}l@{\quad\sim\quad}l@{}}
\quad \mathbf{1}  & \alpha_1\beta_1+\alpha_2\beta_3+\alpha_3\beta_2\\[2mm]
\quad \mathbf{3'}  & \begin{pmatrix} \alpha_2\,\beta_3-\alpha_3\,\beta_2\\
                                    \alpha_1\,\beta_2-\alpha_2\,\beta_1\\
                                    \alpha_3\,\beta_1-\alpha_1\,\beta_3
                    \end{pmatrix}\\[6mm]
\quad \mathbf{5}  & \begin{pmatrix}  2 \alpha_1 \beta_1-\alpha_2 \beta_3-\alpha_3 \beta_2 \\
                                     \sqrt{6}\, \alpha_3 \beta_3 \\
                                     -\sqrt{3}\, \alpha_1 \beta_2-\sqrt{3}\, \alpha_2 \beta_1 \\
                                     -\sqrt{3}\, \alpha_1 \beta_3-\sqrt{3}\, \alpha_3 \beta_1 \\
                                     \sqrt{6}\, \alpha_2 \beta_2
                    \end{pmatrix}
\end{array}\right.
\\[23mm]
\mathbf{3'}&\mathbf{4}&\mathbf{3}\,\oplus\,\mathbf{4}\,\oplus\,\mathbf{5}&
\left\{\begin{array}{@{}l@{\quad\sim\quad}l@{}}
\quad \mathbf{3}   & \begin{pmatrix} -\sqrt{2}\,\alpha_2 \beta_3-\sqrt{2}\,\alpha_3 \beta_2 \\
                                     \sqrt{2}\,\alpha_1 \beta_1+\alpha_2 \beta_4-\alpha_3 \beta_3 \\
                                     \sqrt{2}\,\alpha_1 \beta_4 -\alpha_2 \beta_2+\alpha_3 \beta_1
                     \end{pmatrix}\\[6mm]
\quad \mathbf{4}   & \begin{pmatrix} \alpha_1 \beta_1+\sqrt{2}\,\alpha_3 \beta_3 \\
                                     \alpha_1 \beta_2-\sqrt{2}\,\alpha_3 \beta_4 \\
                                    -\alpha_1 \beta_3+\sqrt{2}\,\alpha_2 \beta_1 \\
                                    -\alpha_1 \beta_4-\sqrt{2}\,\alpha_2 \beta_2
                     \end{pmatrix}\\[9mm]
\quad \mathbf{5}   & \begin{pmatrix}  \sqrt{6}\,\alpha_2 \beta_3-\sqrt{6}\,\alpha_3 \beta_2 \\
                                      \sqrt{2}\,\alpha_1 \beta_1-3\alpha_2 \beta_4-\alpha_3 \beta_3 \\
                                      2\sqrt{2}\,\alpha_1 \beta_2+2 \alpha_3 \beta_4 \\
                                     -2\sqrt{2}\,\alpha_1 \beta_3-2\alpha_2 \beta_1 \\
                                     -\sqrt{2}\,\alpha_1 \beta_4+\alpha_2 \beta_2+3\alpha_3 \beta_1
                     \end{pmatrix}
\end{array}\right.
\\[30mm]
\mathbf{3'}&\mathbf{5}&\mathbf{3}\,\oplus\,\mathbf{3'}\,\oplus\,\mathbf{4}\,\oplus\,\mathbf{5}&
\left\{\begin{array}{@{}l@{\quad\sim\quad}l@{}}
\quad \mathbf{3}   & \begin{pmatrix}
              \sqrt{3}\, \alpha_1 \beta_1+\alpha_2\beta_4+\alpha_3 \beta_3 \\
              \alpha_1 \beta_2-\sqrt{2}\,\alpha_2 \beta_5 -\sqrt{2}\,\alpha_3 \beta_4\\
              \alpha_1 \beta_5-\sqrt{2}\,\alpha_2 \beta_3-\sqrt{2}\,\alpha_3 \beta_2
                     \end{pmatrix}\\[6mm]
\quad \mathbf{3'}   & \begin{pmatrix} 
              -2 \alpha_1 \beta_1+\sqrt{3}\,\alpha_2 \beta_4 +\sqrt{3}\,\alpha_3 \beta_3\\
              \sqrt{3}\,\alpha_1 \beta_3+\alpha_2 \beta_1-\sqrt{6}\,\alpha_3 \beta_5 \\
              \sqrt{3}\,\alpha_1 \beta_4-\sqrt{6}\,\alpha_2 \beta_2+\alpha_3 \beta_1
                     \end{pmatrix}\\[6mm]
\quad \mathbf{4}   & \begin{pmatrix}
              \sqrt{2}\,\alpha_1 \beta_2+3 \alpha_2\beta_5-2\alpha_3 \beta_4 \\
              2\sqrt{2}\,\alpha_1 \beta_3-\sqrt{6}\,\alpha_2 \beta_1+\alpha_3 \beta_5 \\
              -2\sqrt{2}\,\alpha_1 \beta_4-\alpha_2 \beta_2 +\sqrt{6}\,\alpha_3 \beta_1\\
              -\sqrt{2}\,\alpha_1\beta_5+2 \alpha_2\beta_3-3\alpha_3 \beta_2
                     \end{pmatrix}\\[9mm]
\quad \mathbf{5}   & \begin{pmatrix}
              \sqrt{3}\,\alpha_2 \beta_4-\sqrt{3} \alpha_3 \beta_3 \\
              2 \alpha_1 \beta_2+\sqrt{2}\,\alpha_3 \beta_4 \\
              -\alpha_1 \beta_3-\sqrt{3}\,\alpha_2 \beta_1-\sqrt{2}\,\alpha_3 \beta_5 \\
              \alpha_1 \beta_4+\sqrt{2}\,\alpha_2 \beta_2 + \sqrt{3}\,\alpha_3 \beta_1\\
              -2\alpha_1 \beta_5-\sqrt{2}\,\alpha_2 \beta_3
                     \end{pmatrix}
\end{array}\right.
\end{array}
\end{align}
%
%
%
\begin{align}
\begin{array}{@{}cll@{}}
\makecell{\mathbf{4}\,\otimes\,\mathbf{4}{{}\,\,=\,\,{}}\\
\mathbf{1}\,\oplus\,\mathbf{3}\,\oplus\,\mathbf{3'}\\\,\oplus\,\mathbf{4}\,\oplus\,\mathbf{5}}&
\left\{\begin{array}{@{}l@{\quad\sim\quad}l@{}}
\quad \mathbf{1}  & \alpha_1\beta_4+\alpha_2 \beta_3+\alpha_3 \beta_2+\alpha_4 \beta_1 \\[2mm]
\quad \mathbf{3}  & \begin{pmatrix}
              -\alpha_1 \beta_4+\alpha_2\beta_3-\alpha_3\beta_2+\alpha_4 \beta_1\\
              \sqrt{2}\,\alpha_2 \beta_4-\sqrt{2}\,\alpha_4 \beta_2\\
              \sqrt{2}\,\alpha_1 \beta_3-\sqrt{2}\,\alpha_3 \beta_1
                    \end{pmatrix}\\[6mm]
\quad \mathbf{3'}   & \begin{pmatrix}
              \alpha_1 \beta_4 +\alpha_2 \beta_3-\alpha_3 \beta_2 -\alpha_4 \beta_1\\
              \sqrt{2}\,\alpha_3 \beta_4-\sqrt{2}\,\alpha_4 \beta_3 \\
              \sqrt{2}\,\alpha_1 \beta_2-\sqrt{2}\,\alpha_2 \beta_1
                     \end{pmatrix}\\[6mm]
\quad \mathbf{4}   & \begin{pmatrix} 
              \alpha_2 \beta_4+\alpha_3\beta_3+\alpha_4 \beta_2 \\
              \alpha_1 \beta_1+\alpha_3 \beta_4 +\alpha_4 \beta_3\\
              \alpha_1\beta_2+\alpha_2 \beta_1+\alpha_4 \beta_4 \\
              \alpha_1 \beta_3+\alpha_2\beta_2+\alpha_3 \beta_1
                     \end{pmatrix}\\[9mm]
\quad \mathbf{5}   & \begin{pmatrix}
              \sqrt{3}\,\alpha_1 \beta_4-\sqrt{3}\,\alpha_2 \beta_3-\sqrt{3}\,\alpha_3 \beta_2+\sqrt{3}\,\alpha_4 \beta_1 \\
              -\sqrt{2}\,\alpha_2 \beta_4+2\sqrt{2}\,\alpha_3 \beta_3-\sqrt{2}\,\alpha_4 \beta_2 \\
              -2 \sqrt{2}\,\alpha_1 \beta_1+\sqrt{2}\,\alpha_3 \beta_4+\sqrt{2}\,\alpha_4 \beta_3 \\
              \sqrt{2}\,\alpha_1 \beta_2+\sqrt{2}\,\alpha_2 \beta_1-2 \sqrt{2}\,\alpha_4 \beta_4 \\
              -\sqrt{2}\,\alpha_1 \beta_3+2\sqrt{2}\,\alpha_2 \beta_2-\sqrt{2}\,\alpha_3 \beta_1
                     \end{pmatrix}
\end{array}\right.
\\[41mm]
\makecell{\mathbf{4}\,\otimes\,\mathbf{5}{{}\,\,=\,\,{}}\\
\mathbf{3}\,\oplus\,\mathbf{3'}\,\oplus\,\mathbf{4}\\\,\oplus\,\mathbf{5}_1\,\oplus\,\mathbf{5}_2}&
\left\{\begin{array}{@{}l@{\quad\sim\quad}l@{}}
\quad \mathbf{3}   & \begin{pmatrix} 
              2 \sqrt{2}\,\alpha_1\beta_5-\sqrt{2}\,\alpha_2 \beta_4+\sqrt{2}\,\alpha_3 \beta_3-2 \sqrt{2}\,\alpha_4 \beta_2\\
              -\sqrt{6}\,\alpha_1 \beta_1+2 \alpha_2 \beta_5+3 \alpha_3 \beta_4-\alpha_4 \beta_3 \\
              \alpha_1 \beta_4-3 \alpha_2\beta_3-2\alpha_3 \beta_2+\sqrt{6}\,\alpha_4 \beta_1
                     \end{pmatrix}\\[6mm]
\quad \mathbf{3'}   & \begin{pmatrix} 
              \sqrt{2}\,\alpha_1 \beta_5+2\sqrt{2}\,\alpha_2 \beta_4-2\sqrt{2}\,\alpha_3 \beta_3-\sqrt{2}\,\alpha_4 \beta_2 \\
              3\alpha_1 \beta_2-\sqrt{6}\, \alpha_2 \beta_1-\alpha_3 \beta_5+2 \alpha_4\beta_4 \\
              -2 \alpha_1 \beta_3+\alpha_2 \beta_2+\sqrt{6}\,\alpha_3 \beta_1-3 \alpha_4 \beta_5
                     \end{pmatrix}\\[6mm]
\quad \mathbf{4}   & \begin{pmatrix}
              \sqrt{3}\,\alpha_1 \beta_1-\sqrt{2}\,\alpha_2 \beta_5+\sqrt{2}\,\alpha_3 \beta_4-2\sqrt{2}\,\alpha_4 \beta_3 \\
              -\sqrt{2}\,\alpha_1 \beta_2-\sqrt{3}\,\alpha_2 \beta_1+2 \sqrt{2}\,\alpha_3 \beta_5+\sqrt{2}\,\alpha_4 \beta_4 \\
              \sqrt{2}\,\alpha_1 \beta_3+2\sqrt{2}\,\alpha_2 \beta_2-\sqrt{3}\,\alpha_3 \beta_1-\sqrt{2}\,\alpha_4 \beta_5\\
              -2 \sqrt{2}\,\alpha_1 \beta_4+\sqrt{2}\,\alpha_2 \beta_3-\sqrt{2}\,\alpha_3 \beta_2+\sqrt{3}\,\alpha_4 \beta_1 
                     \end{pmatrix}\\[9mm]
\quad \mathbf{5}_1   & \begin{pmatrix}
              \sqrt{2}\,\alpha_1 \beta_5-\sqrt{2}\,\alpha_2 \beta_4-\sqrt{2}\,\alpha_3 \beta_3+\sqrt{2}\,\alpha_4 \beta_2\\
              -\sqrt{2}\,\alpha_1 \beta_1-\sqrt{3}\,\alpha_3 \beta_4 -\sqrt{3}\,\alpha_4 \beta_3 \\
              \sqrt{3}\,\alpha_1 \beta_2+\sqrt{2}\,\alpha_2 \beta_1+\sqrt{3}\,\alpha_3 \beta_5 \\
              \sqrt{3}\,\alpha_2 \beta_2+\sqrt{2}\,\alpha_3 \beta_1+\sqrt{3}\,\alpha_4 \beta_5 \\
              -\sqrt{3}\,\alpha_1 \beta_4-\sqrt{3}\,\alpha_2 \beta_3-\sqrt{2}\,\alpha_4 \beta_1
                     \end{pmatrix}\\[11mm]
\quad \mathbf{5}_2   & \begin{pmatrix} 
              2 \alpha_1\beta_5+4 \alpha_2 \beta_4+4 \alpha_3 \beta_3 +2 \alpha_4 \beta_2\\
              4 \alpha_1 \beta_1+2 \sqrt{6}\,\alpha_2 \beta_5 \\
              -\sqrt{6}\,\alpha_1\beta_2+2 \alpha_2 \beta_1-\sqrt{6}\,\alpha_3 \beta_5 +2 \sqrt{6}\,\alpha_4 \beta_4\\
               2 \sqrt{6}\,\alpha_1 \beta_3-\sqrt{6}\,\alpha_2\beta_2+2 \alpha_3 \beta_1-\sqrt{6}\,\alpha_4 \beta_5 \\
               2 \sqrt{6}\,\alpha_3 \beta_2+4 \alpha_4 \beta_1
                     \end{pmatrix}
\end{array}\right.
\end{array}
\end{align}
%
%
%
\begin{align}
\begin{array}{@{}cll@{}}
\makecell{\mathbf{5}\,\otimes\,\mathbf{5}
{{}\,\,=\,\,{}}
\\ \mathbf{1}\,\oplus\,\mathbf{3}\,\oplus\,\mathbf{3'}\,\oplus\,\mathbf{4}_1\\\,\oplus\,\mathbf{4}_2\,\oplus\,\mathbf{5}_1\,\oplus\,\mathbf{5}_2}&
\left\{\begin{array}{@{}l@{\quad\sim\quad}l@{}}
\quad \mathbf{1}  & \alpha_1\beta_1+\alpha_2\beta_5+\alpha_3\beta_4+\alpha_4\beta_3+\alpha_5\beta_2 \\[2mm]
\quad \mathbf{3}  & \begin{pmatrix} 
     \alpha_2 \beta_5+2\alpha_3 \beta_4-2 \alpha_4 \beta_3-\alpha_5 \beta_2 \\
     -\sqrt{3}\,\alpha_1\beta_2+\sqrt{3}\,\alpha_2 \beta_1+\sqrt{2}\,\alpha_3 \beta_5-\sqrt{2}\,\alpha_5 \beta_3 \\
     \sqrt{3}\,\alpha_1 \beta_5+\sqrt{2}\,\alpha_2 \beta_4-\sqrt{2}\,\alpha_4\beta_2-\sqrt{3}\,\alpha_5 \beta_1
                    \end{pmatrix}\\[6mm]
\quad \mathbf{3'}   & \begin{pmatrix}
     2 \alpha_2 \beta_5-\alpha_3 \beta_4+\alpha_4 \beta_3-2 \alpha_5 \beta_2\\
     \sqrt{3}\,\alpha_1\beta_3-\sqrt{3}\,\alpha_3 \beta_1+\sqrt{2}\,\alpha_4 \beta_5-\sqrt{2}\,\alpha_5 \beta_4 \\
     -\sqrt{3}\,\alpha_1 \beta_4+\sqrt{2}\,\alpha_2 \beta_3-\sqrt{2}\,\alpha_3\beta_2+\sqrt{3}\,\alpha_4 \beta_1
                     \end{pmatrix}\\[6mm]
\quad \mathbf{4}_1   & \begin{pmatrix} 
     3 \sqrt{2}\,\alpha_1 \beta_2+3 \sqrt{2}\,\alpha_2 \beta_1-\sqrt{3}\,\alpha_3\beta_5+4 \sqrt{3}\,\alpha_4\beta_4-\sqrt{3}\,\alpha_5 \beta_3\\
     3 \sqrt{2}\,\alpha_1 \beta_3+4 \sqrt{3}\,\alpha_2 \beta_2+3 \sqrt{2}\,\alpha_3 \beta_1-\sqrt{3}\,\alpha_4\beta_5-\sqrt{3}\, \alpha_5 \beta_4 \\
     3 \sqrt{2}\,\alpha_1 \beta_4-\sqrt{3}\,\alpha_2 \beta_3-\sqrt{3}\,\alpha_3 \beta_2+3 \sqrt{2}\,\alpha_4 \beta_1+4\sqrt{3}\,\alpha_5\beta_5 \\
     3 \sqrt{2}\,\alpha_1\beta_5-\sqrt{3}\,\alpha_2 \beta_4+4 \sqrt{3}\,\alpha_3 \beta_3-\sqrt{3}\,\alpha_4 \beta_2+3\sqrt{2}\,\alpha_5 \beta_1
                     \end{pmatrix}\\[9mm]
\quad \mathbf{4}_2   & \begin{pmatrix}
     \sqrt{2}\,\alpha_1 \beta_2-\sqrt{2}\,\alpha_2 \beta_1+\sqrt{3}\,\alpha_3 \beta_5-\sqrt{3}\,\alpha_5 \beta_3 \\
     -\sqrt{2}\,\alpha_1 \beta_3+\sqrt{2}\,\alpha_3 \beta_1+\sqrt{3}\,\alpha_4 \beta_5-\sqrt{3}\,\alpha_5 \beta_4 \\
     -\sqrt{2}\,\alpha_1 \beta_4-\sqrt{3}\,\alpha_2 \beta_3+\sqrt{3}\,\alpha_3 \beta_2+\sqrt{2}\,\alpha_4 \beta_1\\
     \sqrt{2}\,\alpha_1 \beta_5-\sqrt{3}\,\alpha_2 \beta_4+\sqrt{3}\,\alpha_4 \beta_2-\sqrt{2}\,\alpha_5 \beta_1
                     \end{pmatrix}\\[9mm]
\quad \mathbf{5}_1   & \begin{pmatrix}
     2 \alpha_1 \beta_1+\alpha_2 \beta_5-2 \alpha_3 \beta_4-2 \alpha_4 \beta_3+\alpha_5 \beta_2 \\
     \alpha_1 \beta_2+\alpha_2 \beta_1+\sqrt{6}\,\alpha_3 \beta_5+\sqrt{6}\,\alpha_5 \beta_3 \\
     -2 \alpha_1 \beta_3+\sqrt{6}\,\alpha_2 \beta_2-2 \alpha_3 \beta_1 \\
     -2 \alpha_1 \beta_4-2 \alpha_4 \beta_1+\sqrt{6}\,\alpha_5 \beta_5 \\
     \alpha_1 \beta_5+\sqrt{6}\,\alpha_2 \beta_4+\sqrt{6}\,\alpha_4 \beta_2+\alpha_5 \beta_1
                     \end{pmatrix}\\[11mm]
\quad \mathbf{5}_2   & \begin{pmatrix}
     2 \alpha_1 \beta_1-2 \alpha_2 \beta_5+\alpha_3 \beta_4+\alpha_4\beta_3-2 \alpha_5 \beta_2 \\
     -2 \alpha_1 \beta_2-2 \alpha_2 \beta_1+\sqrt{6}\,\alpha_4 \beta_4 \\
     \alpha_1 \beta_3+\alpha_3 \beta_1+\sqrt{6}\,\alpha_4 \beta_5+\sqrt{6}\,\alpha_5 \beta_4 \\
     \alpha_1 \beta_4+\sqrt{6}\,\alpha_2 \beta_3+\sqrt{6}\,\alpha_3 \beta_2+\alpha_4 \beta_1 \\
     -2 \alpha_1 \beta_5+\sqrt{6}\,\alpha_3 \beta_3-2 \alpha_5 \beta_1
                     \end{pmatrix}
\end{array}\right.
\end{array}
\end{align}

\section{Higher weight Forms and Constraints}
\label{app:constraints}
Through tensor products of $Y_\mathbf{5}$, $Y_\mathbf{3}$ and $Y_\mathbf{3'}$,
one can find, at weight 4, the multiplets:
\begin{equation}
\begin{aligned}
Y^{(4)}_\mathbf{1}    &\,=\, Y_1^2+2 Y_3 Y_4+2 Y_2 Y_5 \,\sim\, \mathbf{1}  \,,\\[2mm]
{Y^{(4)}_\mathbf{1}}' &\,=\,Y_6^2+2 Y_7 Y_8 \,\sim\, \mathbf{1}  \,,\\[2mm]
{Y^{(4)}_\mathbf{1}}'' &\,=\, Y_9^2+2 Y_{10} Y_{11} \,\sim\, \mathbf{1}  \,,
\label{eq:66l}
\end{aligned}
\end{equation}
\begin{equation}
\begin{aligned}
{Y^{(4)}_\mathbf{3}}  &\,=\, 
\left(
\begin{array}{c}
 -2 Y_1 Y_6+\sqrt{3}\, Y_5 Y_7+\sqrt{3}\, Y_2 Y_8 \\
 \sqrt{3}\, Y_2 Y_6+Y_1 Y_7-\sqrt{6}\, Y_3 Y_8 \\
 \sqrt{3}\, Y_5 Y_6-\sqrt{6}\, Y_4 Y_7+Y_1 Y_8 \\
\end{array}
\right)
\,\sim\, \mathbf{3}  \,,\\[2mm]
{Y^{(4)}_\mathbf{3}}'  &\,=\, 
\left(
\begin{array}{c}
 \sqrt{3}\, Y_1 Y_9+Y_4 Y_{10}+Y_3 Y_{11} \\
 Y_2 Y_9-\sqrt{2}\, Y_5 Y_{10}-\sqrt{2}\, Y_4 Y_{11} \\
 Y_5 Y_9-\sqrt{2}\, Y_3 Y_{10}-\sqrt{2}\, Y_2 Y_{11} \\
\end{array}
\right)
\,\sim\, \mathbf{3}  \,,
\end{aligned}
\end{equation}
\begin{equation}
\begin{aligned}
{Y^{(4)}_\mathbf{3'}}  &\,=\,
\left(
\begin{array}{c}
 \sqrt{3}\, Y_1 Y_6+Y_5 Y_7+Y_2 Y_8 \\
 Y_3 Y_6-\sqrt{2}\, Y_2 Y_7-\sqrt{2}\, Y_4 Y_8 \\
 Y_4 Y_6-\sqrt{2}\, Y_3 Y_7-\sqrt{2}\, Y_5 Y_8 \\
\end{array}
\right)
\,\sim\, \mathbf{3'} \,,\\[2mm]
{Y^{(4)}_\mathbf{3'}}'  &\,=\,
\left(
\begin{array}{c}
 -2 Y_1 Y_9+\sqrt{3}\, Y_4 Y_{10}+\sqrt{3}\, Y_3 Y_{11} \\
 \sqrt{3}\, Y_3 Y_9+Y_1 Y_{10}-\sqrt{6}\, Y_5 Y_{11} \\
 \sqrt{3}\, Y_4 Y_9-\sqrt{6}\, Y_2 Y_{10}+Y_1 Y_{11} \\
\end{array}
\right)
\,\sim\, \mathbf{3'}\,,
\end{aligned}
\end{equation}
\begin{equation}
\begin{aligned}
{Y^{(4)}_\mathbf{4}}  &\,=\,
\left(
\begin{array}{c}
 2 Y_4^2+\sqrt{6}\, Y_1 Y_2-Y_3 Y_5 \\
 2 Y_2^2+\sqrt{6}\, Y_1 Y_3-Y_4 Y_5 \\
 2 Y_5^2-Y_2 Y_3+\sqrt{6}\, Y_1 Y_4 \\
 2 Y_3^2-Y_2 Y_4+\sqrt{6}\, Y_1 Y_5 \\
\end{array}
\right)
\,\sim\, \mathbf{4}\,,\\[2mm]
{Y^{(4)}_\mathbf{4}}'  &\,=\,
\left(
\begin{array}{c}
 2 \sqrt{2}\, Y_2 Y_6-\sqrt{6}\, Y_1 Y_7+Y_3 Y_8 \\
 -\sqrt{2}\, Y_3 Y_6+2 Y_2 Y_7-3 Y_4 Y_8 \\
 \sqrt{2}\, Y_4 Y_6+3 Y_3 Y_7-2 Y_5 Y_8 \\
 -2 \sqrt{2}\, Y_5 Y_6-Y_4 Y_7+\sqrt{6}\, Y_1 Y_8 \\
\end{array}
\right)
\,\sim\, \mathbf{4}\,,\\[2mm]
{Y^{(4)}_\mathbf{4}}''  &\,=\,
\left(
\begin{array}{c}
 \sqrt{2}\, Y_7 Y_9+Y_8 Y_{10} \\
 -\sqrt{2}\, Y_6 Y_{10}-Y_8 Y_{11} \\
 -Y_7 Y_{10}-\sqrt{2}\, Y_6 Y_{11} \\
 \sqrt{2}\, Y_8 Y_9+Y_7 Y_{11} \\
\end{array}
\right)
\,\sim\, \mathbf{4}\,,\\[2mm]
{Y^{(4)}_\mathbf{4}}'''  &\,=\,
\left(
\begin{array}{c}
 \sqrt{2}\, Y_2 Y_9+3 Y_5 Y_{10}-2 Y_4 Y_{11} \\
 2 \sqrt{2}\, Y_3 Y_9-\sqrt{6}\, Y_1 Y_{10}+Y_5 Y_{11} \\
 -2 \sqrt{2}\, Y_4 Y_9-Y_2 Y_{10}+\sqrt{6}\, Y_1 Y_{11} \\
 -\sqrt{2}\, Y_5 Y_9+2 Y_3 Y_{10}-3 Y_2 Y_{11} \\
\end{array}
\right)
\,\sim\, \mathbf{4}\,,
\label{eq:66m}
\end{aligned}
\end{equation}
\begin{equation}
\begin{aligned}
{Y^{(4)}_\mathbf{5}}  &\,=\,
\left(
\begin{array}{c}
 \sqrt{2}\, Y_1^2+\sqrt{2}\, Y_3 Y_4-2 \sqrt{2}\, Y_2 Y_5 \\
 \sqrt{3}\, Y_4^2-2 \sqrt{2}\, Y_1 Y_2 \\
 \sqrt{2}\, Y_1 Y_3+2 \sqrt{3}\, Y_4 Y_5 \\
 2 \sqrt{3}\, Y_2 Y_3+\sqrt{2}\, Y_1 Y_4 \\
 \sqrt{3}\, Y_3^2-2 \sqrt{2}\, Y_1 Y_5 \\
\end{array}
\right)
\,\sim\, \mathbf{5}\,,\\[2mm]
{Y^{(4)}_\mathbf{5}}'  &\,=\,
\left(
\begin{array}{c}
 \sqrt{2}\, Y_1^2-2 \sqrt{2}\, Y_3 Y_4+\sqrt{2}\, Y_2 Y_5 \\
 \sqrt{2}\, Y_1 Y_2+2 \sqrt{3}\, Y_3 Y_5 \\
 \sqrt{3}\, Y_2^2-2 \sqrt{2}\, Y_1 Y_3 \\
 \sqrt{3}\, Y_5^2-2 \sqrt{2}\, Y_1 Y_4 \\
 2 \sqrt{3}\, Y_2 Y_4+\sqrt{2}\, Y_1 Y_5 \\
\end{array}
\right)
\,\sim\, \mathbf{5}\,,\\[2mm]
{Y^{(4)}_\mathbf{5}}''  &\,=\,
\left(
\begin{array}{c}
 \sqrt{3}\, Y_5 Y_7-\sqrt{3}\, Y_2 Y_8 \\
 -Y_2 Y_6-\sqrt{3}\, Y_1 Y_7-\sqrt{2}\, Y_3 Y_8 \\
 -2 Y_3 Y_6-\sqrt{2}\, Y_2 Y_7 \\
 2 Y_4 Y_6+\sqrt{2}\, Y_5 Y_8 \\
 Y_5 Y_6+\sqrt{2}\, Y_4 Y_7+\sqrt{3}\, Y_1 Y_8 \\
\end{array}
\right)
\,\sim\, \mathbf{5}\,,\\[2mm]
{Y^{(4)}_\mathbf{5}}'''  &\,=\,
\left(
\begin{array}{c}
 \sqrt{2}\, Y_6^2-\sqrt{2}\, Y_7 Y_8 \\
 -\sqrt{6}\, Y_6 Y_7 \\
 \sqrt{3}\, Y_7^2 \\
 \sqrt{3}\, Y_8^2 \\
 -\sqrt{6}\, Y_6 Y_8 \\
\end{array}
\right)
\,\sim\, \mathbf{5}\,,\\[2mm]
{Y^{(4)}_\mathbf{5}}''''  &\,=\,
\left(
\begin{array}{c}
 \sqrt{3}\, Y_6 Y_9 \\
 Y_7 Y_9-\sqrt{2}\, Y_8 Y_{10} \\
 Y_6 Y_{10}-\sqrt{2}\, Y_8 Y_{11} \\
 Y_6 Y_{11}-\sqrt{2}\, Y_7 Y_{10} \\
 Y_8 Y_9-\sqrt{2}\, Y_7 Y_{11} \\
\end{array}
\right)
\,\sim\, \mathbf{5}\,,\\[2mm]
{Y^{(4)}_\mathbf{5}}'''''  &\,=\,
\left(
\begin{array}{c}
 \sqrt{3}\, Y_4 Y_{10}-\sqrt{3}\, Y_3 Y_{11} \\
 2 Y_2 Y_9+\sqrt{2}\, Y_4 Y_{11} \\
 -Y_3 Y_9-\sqrt{3}\, Y_1 Y_{10}-\sqrt{2}\, Y_5 Y_{11} \\
 Y_4 Y_9+\sqrt{2}\, Y_2 Y_{10}+\sqrt{3}\, Y_1 Y_{11} \\
 -2 Y_5 Y_9-\sqrt{2}\, Y_3 Y_{10} \\
\end{array}
\right)
\,\sim\, \mathbf{5}\,,\\[2mm]
{Y^{(4)}_\mathbf{5}}''''''  &\,=\,
\left(
\begin{array}{c}
 \sqrt{2}\, Y_9^2-\sqrt{2}\, Y_{10} Y_{11} \\
 \sqrt{3}\, Y_{11}^2 \\
 -\sqrt{6}\, Y_9 Y_{10} \\
 -\sqrt{6}\, Y_9 Y_{11} \\
 \sqrt{3}\, Y_{10}^2 \\
\end{array}
\right)
\,\sim\, \mathbf{5}
\,.
\label{eq:66r}
\end{aligned}
\end{equation}

Not all of the above multiplets are expected to be independent.
Indeed, from the $q$-expansions of the $Y_i(\tau)$ given 
in Appendix~\ref{app:qexp} we find 45 constraints between the 66 different $Y_i(\tau)Y_j(\tau)$ products, namely:
\begin{equation}
\begin{aligned}
\frac{3}{5} \left(Y_1^2+2 Y_3 Y_4+2 Y_2 Y_5\right)&=Y_6^2+2 Y_7 Y_8=Y_9^2+2 Y_{10} Y_{11}\,,
\label{eq:constraintsl}
\end{aligned}
\end{equation}
\vskip 2mm
\begin{equation}
\begin{aligned}
-2 \sqrt{3}\, Y_1 Y_6+3 Y_5 Y_7+3 Y_2 Y_8&=2 \sqrt{3}\, Y_1 Y_9+2 Y_4 Y_{10}+2 Y_3 Y_{11}\,,\\
3 Y_2 Y_6+\sqrt{3}\, Y_1 Y_7-3 \sqrt{2}\, Y_3 Y_8&=2 Y_2 Y_9-2 \sqrt{2}\, Y_5 Y_{10}-2 \sqrt{2}\, Y_4 Y_{11}\,,\\
3 Y_5 Y_6-3 \sqrt{2}\, Y_4 Y_7+\sqrt{3}\, Y_1 Y_8&=2 Y_5 Y_9-2 \sqrt{2}\, Y_3 Y_{10}-2 \sqrt{2}\, Y_2 Y_{11}\,,
\end{aligned}
\end{equation}
\vskip 2mm
\begin{equation}
\begin{aligned}
 2 \sqrt{3}\, Y_1 Y_6+2 Y_5 Y_7+2 Y_2 Y_8&=-2 \sqrt{3}\, Y_1 Y_9+3 Y_4 Y_{10}+3 Y_3 Y_{11} \,,\\
 2 Y_3 Y_6-2 \sqrt{2}\, Y_2 Y_7-2 \sqrt{2}\, Y_4 Y_8&=3 Y_3 Y_9+\sqrt{3}\, Y_1 Y_{10}-3 \sqrt{2}\, Y_5 Y_{11} \,,\\
 2 Y_4 Y_6-2 \sqrt{2}\, Y_3 Y_7-2 \sqrt{2}\, Y_5 Y_8&=3 Y_4 Y_9-3 \sqrt{2}\, Y_2 Y_{10}+\sqrt{3}\, Y_1 Y_{11}\,,
\end{aligned}
\end{equation}
\vskip 2mm
\begin{equation}
\begin{aligned}
\sqrt{6}\, Y_1 Y_2 + 2 Y_4^2 - Y_3 Y_5
 &= ({\sqrt{5}}/{7}) \left(
    2\sqrt{2}\, Y_2 Y_6 - \sqrt{6}\, Y_1 Y_7 + Y_3 Y_8 \right) \\
 &= \sqrt{2}\, Y_7 Y_9 + Y_8 Y_{10} \\
 &= -\sqrt{5} \left(
    \sqrt{2}\, Y_2 Y_9+3 Y_5 Y_{10}-2 Y_4 Y_{11} \right)
 \,,\\
\sqrt{6}\, Y_1 Y_3 + 2 Y_2^2 - Y_4 Y_5
 &= -({\sqrt{5}}/{7}) \left(
    \sqrt{2}\, Y_3 Y_6- 2 Y_2 Y_7+3 Y_4 Y_8 \right) \\
 &= -\sqrt{2}\, Y_6 Y_{10}- Y_8 Y_{11}\\
 &= -\sqrt{5} \left(
    2 \sqrt{2}\, Y_3 Y_9 - \sqrt{6}\, Y_1 Y_{10} + Y_5 Y_{11} \right)
 \,,\\
\sqrt{6}\, Y_1 Y_4 + 2 Y_5^2 - Y_2 Y_3
 &= ({\sqrt{5}}/{7}) \left(
    \sqrt{2}\, Y_4 Y_6 + 3 Y_3 Y_7 - 2 Y_5 Y_8 \right) \\
 &= - \sqrt{2}\, Y_6 Y_{11} - Y_7 Y_{10} \\
 &= \sqrt{5} \left(
    2 \sqrt{2}\, Y_4 Y_9 + Y_2 Y_{10} - \sqrt{6}\, Y_1 Y_{11} \right)
 \,,\\
\sqrt{6}\, Y_1 Y_5 + 2 Y_3^2 - Y_2 Y_4 
 &= -({\sqrt{5}}/{7}) \left(
    2 \sqrt{2}\, Y_5 Y_6 - \sqrt{6}\, Y_1 Y_8  + Y_4 Y_7 \right) \\
 &= \sqrt{2}\, Y_8 Y_9 + Y_7 Y_{11} \\
 &= \sqrt{5} \left(
    \sqrt{2}\, Y_5 Y_9 - 2 Y_3 Y_{10} + 3 Y_2 Y_{11} \right)
\,,
\label{eq:constraintsr}
\end{aligned}
\end{equation}
\vskip 2mm
\begin{equation}
\begin{aligned}
 Y_2 Y_8-Y_5 Y_7 &= 2 Y_4 Y_{10}-2 Y_3 Y_{11} \,,\\
 Y_2 Y_6+\sqrt{3}\, Y_1 Y_7+\sqrt{2}\, Y_3 Y_8 &= 4 Y_2 Y_9+2 \sqrt{2}\, Y_4 Y_{11} \,,\\
 \sqrt{2}\, Y_3 Y_6+Y_2 Y_7 &= -\sqrt{2}\, Y_3 Y_9-\sqrt{6}\, Y_1 Y_{10}-2 Y_5 Y_{11} \,,\\
 -\sqrt{2}\, Y_4 Y_6-Y_5 Y_8 &= \sqrt{2}\, Y_4 Y_9+2 Y_2 Y_{10}+\sqrt{6}\, Y_1 Y_{11} \,,\\
 Y_5 Y_6+\sqrt{2}\, Y_4 Y_7+\sqrt{3}\, Y_1 Y_8 &= 4 Y_5 Y_9 + 2 \sqrt{2}\, Y_3 Y_{10} \,,
\label{eq:constraintsl2}
\end{aligned}
\end{equation}
\vskip 2mm
\begin{equation}
\begin{aligned}
 4 Y_1^2+4 Y_3 Y_4-8 Y_2 Y_5- 3 \sqrt{5}\, Y_5 Y_7 - 3 \sqrt{5}\, Y_2 Y_8 &= 4 Y_1^2-8 Y_3 Y_4+4 Y_2 Y_5
\,,\\
 2 \sqrt{6}\, Y_4^2-8 Y_1 Y_2-\sqrt{15}\, Y_2 Y_6- 3 \sqrt{5}\, Y_1 Y_7 -\sqrt{30}\, Y_3 Y_8 &= 4 Y_1 Y_2+4 \sqrt{6}\, Y_3 Y_5
\,,\\
 2 \sqrt{2}\, Y_1 Y_3 - \sqrt{30}\, Y_3 Y_6+4 \sqrt{3}\, Y_4 Y_5 -\sqrt{15}\, Y_2 Y_7 &= 2 \sqrt{3}\, Y_2^2-4 \sqrt{2}\, Y_1 Y_3
\,,\\
 4 \sqrt{3}\, Y_2 Y_3+2 \sqrt{2}\, Y_1 Y_4 + \sqrt{30}\, Y_4 Y_6
 + \sqrt{15}\, Y_5 Y_8 &= 2 \sqrt{3}\, Y_5^2-4 \sqrt{2}\, Y_1 Y_4
\,,\\
 2 \sqrt{6}\, Y_3^2-8 Y_1 Y_5 + \sqrt{15}\, Y_5 Y_6
 + \sqrt{30}\, Y_4 Y_7-15 Y_1 Y_8 &= 4 \sqrt{6}\, Y_2 Y_4+4 Y_1 Y_5
\,,
\end{aligned}
\end{equation}
\vskip 2mm
\begin{equation}
\begin{aligned}
 12 Y_1^2+12 Y_3 Y_4-24 Y_2 Y_5 + 21 \sqrt{5}\, Y_5 Y_7 - 21 \sqrt{5}\, Y_2 Y_8 &= 20 Y_6^2- 20 Y_7 Y_8
\,,\\
 6 \sqrt{2}\, Y_4^2-8 \sqrt{3}\, Y_1 Y_2 - 7 \sqrt{5}\, Y_2 Y_6
 - 7 \sqrt{15}\, Y_1 Y_7 - 7 \sqrt{10}\, Y_3 Y_8 &= - 20 Y_6 Y_7 
\,,\\
 2 \sqrt{6}\, Y_1 Y_3 - 7 \sqrt{10}\, Y_3 Y_6+12 Y_4 Y_5 - 7 \sqrt{5}\, Y_2 Y_7 &= 10 Y_7^2 
\,,\\
 12 Y_2 Y_3+2 \sqrt{6}\, Y_1 Y_4 + 7 \sqrt{10}\, Y_4 Y_6 + 7 \sqrt{5}\, Y_5 Y_8 &= 10 Y_8^2
\,,\\
 6 \sqrt{2}\, Y_3^2-8 \sqrt{3}\, Y_1 Y_5 + 7 \sqrt{5}\, Y_5 Y_6
 + 7  \sqrt{10}\, Y_4 Y_7 + 7 \sqrt{15}\, Y_1 Y_8 &= -20 Y_6 Y_8 \,,
\end{aligned}
\end{equation}
\vskip 2mm
\begin{equation}
\begin{aligned}
 12 Y_1^2+12 Y_3 Y_4-24 Y_2 Y_5 + \sqrt{5}\, Y_5 Y_7 - \sqrt{5}\, Y_2 Y_8 &= -20 Y_6 Y_9
\,,\\
 18 \sqrt{2}\, Y_4^2-24 \sqrt{3}\, Y_1 Y_2 - \sqrt{5}\, Y_2 Y_6
 - \sqrt{15}\, Y_1 Y_7 - \sqrt{10}\, Y_3 Y_8 &= 20 \sqrt{2}\, Y_8 Y_{10}-20 Y_7 Y_9
\,,\\
 6 \sqrt{6}\, Y_1 Y_3 - \sqrt{10}\, Y_3 Y_6 + 36 Y_4 Y_5 - \sqrt{5} Y_2 Y_7 &= 20 Y_8 Y_{11}-10 \sqrt{2}\, Y_6 Y_{10}
\,,\\
 36 Y_2 Y_3+6 \sqrt{6}\, Y_1 Y_4 + \sqrt{10}\, Y_4 Y_6 + \sqrt{5}\, Y_5 Y_8 &= 20 Y_7 Y_{10}-10 \sqrt{2}\, Y_6 Y_{11} \,,\\
 18 \sqrt{2}\, Y_3^2-24 \sqrt{3}\, Y_1 Y_5 + \sqrt{5}\, Y_5 Y_6
 + \sqrt{10}\, Y_4 Y_7 + \sqrt{15}\, Y_1 Y_8 &= 20 \sqrt{2}\, Y_7 Y_{11}-20 Y_8 Y_9 \,,
\end{aligned}
\end{equation}
\vskip 2mm
\begin{equation}
\begin{aligned}
 6 Y_1^2+6 Y_3 Y_4-12 Y_2 Y_5+3\sqrt{5}\, Y_5 Y_7 - 3\sqrt{5}\, Y_2 Y_8 &= 10 Y_9^2-10 Y_{10} Y_{11}
\,,\\
 3 \sqrt{2}\, Y_4^2-4 \sqrt{3}\, Y_1 Y_2- \sqrt{5}\, Y_2 Y_6 - \sqrt{15}\, Y_1 Y_7 - \sqrt{10}\, Y_3 Y_8 &= 5 \sqrt{2}\, Y_{11}^2
\,,\\
 \sqrt{6}\, Y_1 Y_3 - \sqrt{10}\, Y_3 Y_6+6 Y_4 Y_5 - \sqrt{5}\,Y_2 Y_7 &= -5 \sqrt{2}\, Y_9 Y_{10}
\,,\\
 6 Y_2 Y_3+\sqrt{6}\, Y_1 Y_4 + \sqrt{10}\, Y_4 Y_6 + \sqrt{5}\, Y_5 Y_8 &= -5 \sqrt{2}\, Y_9 Y_{11}
\,,\\
 3 \sqrt{2}\, Y_3^2-4 \sqrt{3}\, Y_1 Y_5 + \sqrt{5}\, Y_5 Y_6
 + \sqrt{10}\, Y_4 Y_7 + \sqrt{15}\, Y_1 Y_8 &= 5 \sqrt{2}\, Y_{10}^2
\,.
\label{eq:constraintsr2}
\end{aligned}
\end{equation}
The 20 constraints in eqs.~\eqref{eq:constraintsl}\,--\,\eqref{eq:constraintsr}
imply that the primed multiplets 
${Y_{\mathbf{r}}^{(4)}}^{\prime\ldots\prime}$
in eqs.~\eqref{eq:66l}\,--\,\eqref{eq:66m} 
are proportional among themselves and to the corresponding unprimed ones,
${Y_{\mathbf{r}}^{(4)}}$.
Therefore, for $\mathbf{r} = \mathbf{1}$, $\mathbf{3}$, $\mathbf{3'}$, $\mathbf{4}$, 
only unprimed multiplets are kept
in our discussion.
In what concerns the quintets, it follows instead from
the 25 constraints in eqs.~\eqref{eq:constraintsl2}\,--\,\eqref{eq:constraintsr2}
that there are two independent multiplets
out of the 7 given in eq.~\eqref{eq:66r}, which we take to be $Y^{(4)}_{\mathbf{5},1} \equiv Y^{(4)}_{\mathbf{5}}$ and $Y^{(4)}_{\mathbf{5},2} \equiv {Y^{(4)}_{\mathbf{5}}}''$, cf.~eq.~\eqref{eq:21}.

We also collect here the linearly independent modular multiplets arising at weights 6, 8 and 10. The linear space of modular forms of weight $k$ (and level $N=5$, corresponding to $\Gamma_5\simeq A_5$) has dimension $5k+1$. At weight $k=6$, one finds
\begin{equation}
\begin{aligned}
Y_\mathbf{1}^{(6)} &= 
3 \sqrt{3}\, \left(Y_2 Y_3^2+Y_4^2 Y_5\right)+\sqrt{2}\, Y_1 \left(Y_1^2+3 Y_3 Y_4-6 Y_2 Y_5\right)
\,\sim\, \mathbf{1}  \,,\\[2mm]
Y_{\mathbf{3},1}^{(6)} &= 
\left(Y_1^2+2 Y_3 Y_4+2 Y_2 Y_5\right) 
 \left(\begin{array}{c}
 Y_6\\
 Y_7\\
 Y_8
 \end{array}\right)
\sim\, \mathbf{3}  \,,\\[2mm]
Y_{\mathbf{3},2}^{(6)} &= 
 \left(\begin{array}{c}
  \left(Y_5 Y_6-\sqrt{2}\, Y_4 Y_7\right) Y_7
+\left(\sqrt{2}\, Y_3 Y_8-Y_2 Y_6\right) Y_8\\
 \left(\sqrt{3}\, Y_1 Y_6-Y_5 Y_7\right) Y_7 
-\sqrt{2}\, Y_3 Y_6 Y_8
+\left(Y_6^2-Y_7 Y_8\right) Y_2\\
\left(Y_2 Y_8 -\sqrt{3}\, Y_1 Y_6\right) Y_8
+\sqrt{2}\, Y_4 Y_6 Y_7
-\left(Y_6^2-Y_7 Y_8\right) Y_5
\end{array}\right)
\sim\, \mathbf{3}  \,,\\[2mm]
Y_{\mathbf{3'},1}^{(6)} &=
\left(Y_1^2+2 Y_3 Y_4+2 Y_2 Y_5\right) 
 \left(\begin{array}{c}
  Y_9 \\
 Y_{10} \\
 Y_{11}
 \end{array}\right)
\sim\, \mathbf{3'}  \,,\\[2mm]
Y_{\mathbf{3'},2}^{(6)} &= 
 \left(\begin{array}{c}
\left(Y_4 Y_6-\sqrt{2}\, Y_3 Y_7-\sqrt{2}\, Y_5 Y_8\right) Y_{10}-\left(Y_3 Y_6-\sqrt{2}\, Y_2 Y_7-\sqrt{2}\, Y_4 Y_8\right) Y_{11} \\
\left(Y_3 Y_6-\sqrt{2}\, Y_2 Y_7-\sqrt{2}\, Y_4 Y_8\right) Y_9-\left(\sqrt{3}\, Y_1 Y_6+Y_5 Y_7+Y_2 Y_8\right) Y_{10} \\
\left(\sqrt{3}\, Y_1 Y_6+Y_5 Y_7+Y_2 Y_8\right) Y_{11}-\left(Y_4 Y_6-\sqrt{2}\, Y_3 Y_7-\sqrt{2}\, Y_5 Y_8\right) Y_9
 \end{array}\right)
\sim\, \mathbf{3'}  \,,\\[2mm]
Y_{\mathbf{4},1}^{(6)} &= 
 \left(\begin{array}{c}
\sqrt{2} \left(\sqrt{6}\, Y_3 Y_8 -\sqrt{3}\, Y_2 Y_6-Y_1 Y_7\right) Y_9-\left(\sqrt{3}\, Y_5 Y_6-\sqrt{6}\, Y_4 Y_7+Y_1 Y_8\right) Y_{10} \\
\left(\sqrt{3}\, Y_5 Y_6-\sqrt{6}\, Y_4 Y_7+Y_1 Y_8\right) Y_{11}+\sqrt{2} \left(\sqrt{3}\, Y_5 Y_7-2 Y_1 Y_6+\sqrt{3}\, Y_2 Y_8\right) Y_{10} \\
\left(\sqrt{3}\, Y_2 Y_6+Y_1 Y_7-\sqrt{6}\, Y_3 Y_8\right) Y_{10}+\sqrt{2} \left(\sqrt{3}\, Y_5 Y_7 -2 Y_1 Y_6+\sqrt{3}\, Y_2 Y_8\right) Y_{11} \\
\sqrt{2} \left(\sqrt{6}\, Y_4 Y_7-\sqrt{3}\, Y_5 Y_6-Y_1 Y_8\right) Y_9-\left(\sqrt{3}\, Y_2 Y_6+Y_1 Y_7-\sqrt{6}\, Y_3 Y_8\right) Y_{11}
 \end{array}\right)
\sim\, \mathbf{4}  \,,\\[2mm]
Y_{\mathbf{4},2}^{(6)} &= 
 \left(\begin{array}{c}
 \sqrt{2}\left(\sqrt{3}\, Y_1 Y_6+Y_5 Y_7\right) Y_7 + \left(Y_3 Y_6-\sqrt{2}\, Y_4 Y_8\right)Y_8\\
\sqrt{2}\left(\sqrt{2}\, Y_2 Y_7-Y_3 Y_6\right)Y_6 +\left(Y_4 Y_6+\sqrt{2}\, Y_3 Y_7+\sqrt{2}\, Y_5 Y_8\right) Y_8 \\
\sqrt{2} \left(\sqrt{2}\, Y_5 Y_8-Y_4 Y_6\right)Y_6+\left(Y_3 Y_6+\sqrt{2}\, Y_2 Y_7+\sqrt{2}\, Y_4 Y_8\right) Y_7 \\
\sqrt{2} \left(\sqrt{3}\, Y_1 Y_6+Y_2 Y_8\right) Y_8 + \left(Y_4 Y_6-\sqrt{2}\, Y_3 Y_7\right) Y_7 
 \end{array}\right)
\sim\, \mathbf{4}  \,,\\[2mm]
Y_{\mathbf{5},1}^{(6)} &= 
\left(Y_1^2+2 Y_3 Y_4+2 Y_2 Y_5\right)
 \left(\begin{array}{c}
 Y_1 \\
 Y_2 \\
 Y_3 \\
 Y_4 \\
 Y_5
 \end{array}\right)
\sim\, \mathbf{5}  \,,\\[2mm]
Y_{\mathbf{5},2}^{(6)} &= 
 \left(\begin{array}{c}
\sqrt{3} \left(\sqrt{3}\, Y_1 Y_6+Y_5 Y_7+Y_2 Y_8\right) Y_6\\
\left(Y_5 Y_7+\sqrt{3}\, Y_1 Y_6\right) Y_7+\left(3 Y_2 Y_7+2 Y_4 Y_8 -\sqrt{2}\, Y_3 Y_6\right) Y_8\\
\left(Y_3 Y_6-\sqrt{2}\, Y_2 Y_7\right) Y_6 +2 \left(Y_5 Y_8 + Y_3 Y_7-\sqrt{2}\, Y_4 Y_6\right) Y_8\\
\left(Y_4 Y_6-\sqrt{2}\, Y_5 Y_8\right) Y_6 
+ 2 \left(Y_2 Y_7+Y_4 Y_8 -\sqrt{2}\, Y_3 Y_6\right) Y_7\\
\left(Y_2 Y_8 + \sqrt{3}\, Y_1 Y_6\right) Y_8 + \left(3 Y_5 Y_8 + 2 Y_3 Y_7-\sqrt{2}\, Y_4 Y_6\right) Y_7
 \end{array}\right)
\sim\, \mathbf{5}  \,,
\end{aligned} \nonumber
\end{equation}
%
corresponding to a total dimension of 31. 

At weight $k=8$, one has
\begin{equation}
\begin{aligned}
Y_\mathbf{1}^{(8)} &= 
\left(Y_1^2+2 Y_3 Y_4+2 Y_2 Y_5\right) Y_6^2+2 \left(Y_1^2+2 Y_3 Y_4+2 Y_2 Y_5\right) Y_7 Y_8\,\sim\, \mathbf{1}  \,,\\[2mm]
Y_{\mathbf{3},1}^{(8)} &= 
\left(3 \sqrt{3}\, \left(Y_2 Y_3^2+Y_4^2 Y_5\right)+\sqrt{2}\, Y_1 \left(Y_1^2+3 Y_3 Y_4-6 Y_2 Y_5\right)\right) 
 \left(\begin{array}{c}
Y_6\\
Y_7\\
Y_8
 \end{array}\right)
\sim\, \mathbf{3}  \,,\\[2mm]
Y_{\mathbf{3},2}^{(8)} &= 
\left(Y_1^2+2 Y_3 Y_4+2 Y_2 Y_5\right)
 \left(\begin{array}{c}
-2 Y_1 Y_6 + \sqrt{3}\,Y_5 Y_7+ \sqrt{3}\,Y_2 Y_8\\
\sqrt{3}\, Y_2 Y_6+Y_1 Y_7-\sqrt{6}\, Y_3 Y_8\\
\sqrt{3}\, Y_5 Y_6-\sqrt{6}\, Y_4 Y_7+Y_1 Y_8
 \end{array}\right)
\sim\, \mathbf{3}  \,,\\[2mm]
Y_{\mathbf{3'},1}^{(8)} &= 
\left(3 \sqrt{3}\, \left(Y_2 Y_3^2+Y_4^2 Y_5\right)+\sqrt{2}\, Y_1 \left(Y_1^2+3 Y_3 Y_4-6 Y_2 Y_5\right)\right) 
 \left(\begin{array}{c}
Y_9\\
Y_{10}\\
Y_{11}
 \end{array}\right)
\sim\, \mathbf{3'}  \,,\\[2mm]
Y_{\mathbf{3'},2}^{(8)} &= 
\left(Y_1^2+2 Y_3 Y_4+2 Y_2 Y_5\right) 
 \left(\begin{array}{c}
\sqrt{3}\, Y_1 Y_6+Y_5 Y_7+Y_2 Y_8\\
Y_3 Y_6 -\sqrt{2}\, Y_2 Y_7 -\sqrt{2}\, Y_4 Y_8\\
Y_4 Y_6 -\sqrt{2}\, Y_3 Y_7 -\sqrt{2}\, Y_5 Y_8
 \end{array}\right)
\sim\, \mathbf{3'}  \,,\\[2mm]
Y_{\mathbf{4},1}^{(8)} &= 
 \left(Y_1^2+2 Y_3 Y_4+2 Y_2 Y_5\right)
 \left(\begin{array}{c}
 \sqrt{2}\, Y_7 Y_9+Y_8 Y_{10}\\
-\sqrt{2}\, Y_6 Y_{10}-Y_8 Y_{11}\\
-Y_7 Y_{10}-\sqrt{2}\, Y_6 Y_{11}\\
\sqrt{2}\, Y_8 Y_9+Y_7 Y_{11}
 \end{array}\right)
\sim\, \mathbf{4}  \,,\\[2mm]
Y_{\mathbf{4},2}^{(8)} &= 
 \left(\begin{array}{l}
 \sqrt{3}\, Y_1 \left(\sqrt{6}\, Y_1 \left(Y_4^2+Y_3 Y_5\right)+4 Y_5 \left(Y_2^2+Y_4 Y_5\right)-6 Y_1^2 Y_2-Y_3^3\right)\\
\qquad+\sqrt{2} \left(12 Y_2^2 Y_3^2-Y_3 Y_4 \left(Y_4^2+Y_3 Y_5\right)+2 Y_2 \left(Y_3 Y_5-5 Y_4^2\right) Y_5\right) \\[2mm]
 \sqrt{3}\, Y_1 \left(\sqrt{6}\, Y_1 \left(2 Y_2^2+3 Y_4 Y_5\right)-3 Y_4 Y_3^2-4 \left(2 Y_4^2+Y_3 Y_5\right) Y_2\right)\\
\qquad-\sqrt{2} \left(3 Y_2 Y_3^3 +4 Y_2^2 Y_3 Y_4 -3 Y_4^4
+4 \left(Y_2^3-2 Y_3 Y_4^2\right) Y_5 +10 Y_2 Y_5^2 Y_4 \right) \\[2mm]
 \sqrt{3}\, Y_1 \left(\sqrt{6}\, Y_1 \left(2 Y_5^2+3 Y_2 Y_3\right)-3 Y_3 Y_4^2-4 \left(2 Y_3^2+Y_2 Y_4\right) Y_5\right)\\
\qquad-\sqrt{2} \left(3 Y_4^3 Y_5+4 Y_4 Y_5^2 Y_3-3 Y_3^4 +4 Y_2 \left(Y_5^3-2 Y_4 Y_3^2\right)+10 Y_2^2 Y_5 Y_3 \right) \\[2mm]
 \sqrt{3}\, Y_1 \left(\sqrt{6}\, Y_1 \left(Y_3^2+Y_2 Y_4\right) +4 Y_2 \left(Y_5^2+Y_2 Y_3\right)-6 Y_1^2 Y_5-Y_4^3\right)\\
\qquad+\sqrt{2} \left(12 Y_4^2 Y_5^2-Y_3 Y_4 \left(Y_3^2+Y_2 Y_4\right)+2 Y_2 \left(Y_4 Y_2-5 Y_3^2\right) Y_5\right)
 \end{array}\right)
\sim\, \mathbf{4}  \,,\\[2mm]
Y_{\mathbf{5},1}^{(8)} &= 
\left(Y_1^2+2 Y_3 Y_4+2 Y_2 Y_5\right)
 \left(\begin{array}{c}
 \sqrt{2} \left(Y_1^2+Y_3 Y_4-2 Y_2 Y_5\right)\\
 \sqrt{3}\, Y_4^2-2 \sqrt{2}\, Y_1 Y_2\\
 \sqrt{2} \left(Y_1 Y_3+\sqrt{6}\, Y_4 Y_5\right)\\
 \sqrt{2} \left(Y_1 Y_4+\sqrt{6}\, Y_2 Y_3\right)\\
 \sqrt{3}\, Y_3^2-2 \sqrt{2}\, Y_1 Y_5
 \end{array}\right)
\sim\, \mathbf{5}  \,,\\[2mm]
Y_{\mathbf{5},2}^{(8)} &= 
\left(Y_1^2+2 Y_3 Y_4+2 Y_2 Y_5\right)
 \left(\begin{array}{c}
 \sqrt{2} \left(Y_1^2+Y_2 Y_5-2 Y_3 Y_4\right)\\
 \sqrt{2}\, \left(Y_1 Y_2+\sqrt{6}\, Y_3 Y_5\right)\\
 \sqrt{3}\, Y_2^2-2 \sqrt{2}\, Y_1 Y_3\\
 \sqrt{3}\, Y_5^2-2 \sqrt{2}\, Y_1 Y_4\\
 \sqrt{2} \left(Y_1 Y_5 + \sqrt{6}\, Y_2 Y_4\right)
 \end{array}\right)
\sim\, \mathbf{5}  \,,
\end{aligned} \nonumber
\end{equation}
\begin{equation}
\begin{aligned}
Y_{\mathbf{5},3}^{(8)} &= 
\left(3 \sqrt{3}\, \left(Y_2 Y_3^2+Y_4^2 Y_5\right)+\sqrt{2}\, Y_1 \left(Y_1^2+3 Y_3 Y_4-6 Y_2 Y_5\right)\right) 
 \left(\begin{array}{c}
 Y_1 \\
 Y_2 \\
 Y_3 \\
 Y_4 \\
 Y_5
 \end{array}\right)
\sim\, \mathbf{5}  \,,\\[2mm]
Y_{\mathbf{5},4}^{(8)} &= 
 \left(\begin{array}{l}
  Y_2 \left(6 \sqrt{6}\, Y_5 Y_1^2-3 Y_3^2 Y_1-2 \sqrt{6}\, Y_2^2 Y_3+\sqrt{6}\, Y_4 \left(Y_4^2+2 Y_3 Y_5\right)
  \right. \\ \left.
  -2 Y_2 \left(2 \sqrt{6}\, Y_5^2+3 Y_1 Y_4\right)\right)
  +Y_5 \left(\sqrt{6}\, Y_3^3-6 Y_1 Y_5 Y_3-Y_4 \left(2 \sqrt{6}\, Y_5^2+3 Y_1 Y_4\right)\right) \\[2mm]
 Y_1 \left(3 Y_1 \left(Y_4^2+4 Y_3 Y_5\right) + \sqrt{6}\left(2 Y_5 \left(Y_2^2+Y_4 Y_5\right)-2 Y_3^3-3 Y_2 Y_4 Y_3\right)\right)\\
\qquad-Y_2 \left(3 \sqrt{6}\, Y_1^3+9 Y_2 Y_3^2+Y_5\left(4 Y_3 Y_5-5 Y_4^2\right) \right)+2 Y_3 Y_4 \left(Y_4^2+Y_3 Y_5\right) \\[2mm]
 Y_2 \left(\sqrt{6}\,Y_1\left(Y_4^2+2 Y_3 Y_5\right) -6 Y_2 Y_1^2-2 \left(6 Y_3^3+Y_2 Y_4 Y_3\right)\right)\\
\qquad+2Y_5 \left(5 Y_3 Y_4^2+2 Y_2 \left(Y_2^2+Y_4 Y_5\right)\right) \\[2mm]
Y_5 \left(\sqrt{6}\, Y_1\left(Y_3^2+2 Y_2 Y_4\right) -6 Y_5 Y_1^2-2 \left(6 Y_4^3+Y_3 Y_5 Y_4\right)\right)\\
\qquad+2 Y_2 \left(5 Y_4 Y_3^2+2 Y_5\left(Y_5^2+Y_2 Y_3\right) \right)
 \\[2mm]
Y_1 \left(3 Y_1 \left(Y_3^2+4 Y_2 Y_4\right)+\sqrt{6}\left(2 Y_2 \left(Y_5^2+Y_2 Y_3\right)-2 Y_4^3-3 Y_3 Y_5 Y_4\right)\right) \\
\qquad-Y_5 \left(3 \sqrt{6}\, Y_1^3+9 Y_4^2 Y_5+ Y_2\left(4 Y_2 Y_4-5 Y_3^2\right)\right)
+  2 Y_3 Y_4 \left(Y_3^2+Y_2 Y_4\right)
 \end{array}\right)
\sim\, \mathbf{5}  \,,
\end{aligned} \nonumber
\end{equation}
%
corresponding to a total dimension of 41.

Finally, at weight $k=10$, one has
\begin{equation}
\begin{aligned}
Y_\mathbf{1}^{(10)} &= 
3 \sqrt{3} \left(Y_2 Y_3^2+Y_4^2 Y_5\right) + \sqrt{2}\, Y_1^3+3 \sqrt{2} \left(Y_3 Y_4-2 Y_2 Y_5\right) Y_1
\,\sim\, \mathbf{1}  \,,\\[2mm]
Y_{\mathbf{3},1}^{(10)} &= 
\left(Y_1^2+2 Y_3 Y_4+2 Y_2 Y_5\right)^2
 \left(\begin{array}{c}
 Y_6 \\ 
 Y_7 \\ 
 Y_8
 \end{array}\right)
\sim\, \mathbf{3}  \,,\\[2mm]
Y_{\mathbf{3},2}^{(10)} &= 
\left(Y_1^2+2 Y_3 Y_4+2 Y_2 Y_5\right)
\\&\quad \times
 \left(\begin{array}{l}
\left(3 Y_1^2+2 Y_3 Y_4\right) Y_6-\left(\sqrt{2}\, Y_3^2+\sqrt{2}\, Y_2 Y_4-\sqrt{3}\, Y_1 Y_5\right) Y_7\\
\quad+\left(\sqrt{3}\, Y_1 Y_2-\sqrt{2}\, \left(Y_4^2+Y_3 Y_5\right)\right) Y_8\\[2mm]
\left(\sqrt{3}\, Y_1 Y_2-\sqrt{2}\, \left(Y_4^2+Y_3 Y_5\right)\right) Y_6+\left(2 Y_3 Y_4+3 Y_2 Y_5\right) Y_7\\
\quad+\left(Y_2^2+4 Y_4 Y_5\right) Y_8\\[2mm]
-\left(\sqrt{2}\, Y_3^2+\sqrt{2}\, Y_2 Y_4-\sqrt{3}\, Y_1 Y_5\right) Y_6+\left(Y_5^2+4 Y_2 Y_3\right) Y_7\\
\quad+\left(2 Y_3 Y_4+3 Y_2 Y_5\right) Y_8
 \end{array}\right)
\sim\, \mathbf{3}  \,,\\[2mm]
Y_{\mathbf{3},3}^{(10)} &= 
\left(3 \sqrt{3}\, \left(Y_2 Y_3^2+Y_4^2 Y_5\right)+\sqrt{2}\, Y_1 \left(Y_1^2+3 Y_3 Y_4-6 Y_2 Y_5\right)\right)\\
&\quad \times
 \left(\begin{array}{c}
 \sqrt{3} \left(Y_5 Y_7+Y_2 Y_8\right)-2 Y_1 Y_6\\
 \sqrt{3}\, Y_2 Y_6+Y_1 Y_7-\sqrt{6}\, Y_3 Y_8\\
 \sqrt{3}\, Y_5 Y_6-\sqrt{6}\, Y_4 Y_7+Y_1 Y_8
 \end{array}\right)
\sim\, \mathbf{3}  \,,
\end{aligned} \nonumber
\end{equation}
%
\begin{equation}
\begin{aligned}
Y_{\mathbf{3'},1}^{(10)} &= 
\left(Y_1^2+2 Y_3 Y_4+2 Y_2 Y_5\right)^2
 \left(\begin{array}{c}
 Y_9 \\ 
 Y_{10} \\ 
 Y_{11}
 \end{array}\right)
\sim\, \mathbf{3'}  \,,\\[2mm]
Y_{\mathbf{3'},2}^{(10)} &= 
\left(Y_1^2+2 Y_3 Y_4+2 Y_2 Y_5\right)
\\&\quad \times
\left(\begin{array}{l}
-2 \left(Y_1^2+Y_3 Y_4-2 Y_2 Y_5\right) Y_9+\left(3 \sqrt{2}\, Y_2 Y_3+\sqrt{3}\, Y_1 Y_4\right) Y_{10}\\
\quad+\left(\sqrt{3}\, Y_1 Y_3+3 \sqrt{2}\, Y_4 Y_5\right) Y_{11}\\[2mm]
\left(\sqrt{3}\, Y_1 Y_3+3 \sqrt{2}\, Y_4 Y_5\right) Y_9+\left(Y_1^2+Y_3 Y_4-2 Y_2 Y_5\right) Y_{10}\\
\quad+\left(2 \sqrt{6}\, Y_1 Y_5-3 Y_3^2\right) Y_{11}\\[2mm]
\left(3 \sqrt{2}\, Y_2 Y_3+\sqrt{3}\, Y_1 Y_4\right) Y_9+\left(2 \sqrt{6}\, Y_1 Y_2-3 Y_4^2\right) Y_{10}\\
\quad+\left(Y_1^2+Y_3 Y_4-2 Y_2 Y_5\right) Y_{11}
 \end{array}\right)
\sim\, \mathbf{3'}  \,,\\[2mm]
Y_{\mathbf{3'},3}^{(10)} &= 
\left(3 \sqrt{3}\, \left(Y_2 Y_3^2+Y_4^2 Y_5\right)+\sqrt{2}\, Y_1 \left(Y_1^2+3 Y_3 Y_4-6 Y_2 Y_5\right)\right)\\
&\quad \times
 \left(\begin{array}{c}
 \sqrt{3} \left(Y_4 Y_{10}+Y_3 Y_{11}\right)-2 Y_1 Y_9\\
 \sqrt{3}\, Y_3 Y_9+Y_1 Y_{10}-\sqrt{6}\, Y_5 Y_{11}\\
 \sqrt{3}\, Y_4 Y_9-\sqrt{6}\, Y_2 Y_{10}+Y_1 Y_{11}
 \end{array}\right)
\sim\, \mathbf{3'}  \,,\\[2mm]
Y_{\mathbf{4},1}^{(10)} &= 
\left(Y_1^2+2 Y_3 Y_4+2 Y_2 Y_5\right)\\
&\quad \times
 \left(\begin{array}{c}
 -\left(Y_3 Y_5-2 Y_4^2-\sqrt{6}\, Y_1 Y_2\right) Y_6+\left(\sqrt{2}\left(Y_4 Y_5-2 Y_2^2 \right )-2 \sqrt{3}\, Y_1 Y_3\right) Y_8
\\
\left(Y_4 Y_5-2 Y_2^2-\sqrt{6}\, Y_1 Y_3\right) Y_6+\left(\sqrt{2} \left(Y_3 Y_5-2 Y_4^2\right)-2 \sqrt{3}\, Y_1 Y_2\right) Y_7\\
-\left(Y_2 Y_3-2 Y_5^2-\sqrt{6}\, Y_1 Y_4\right) Y_6
-\left(\sqrt{2}\left(Y_2 Y_4-2 \, Y_3^2\right)-2 \sqrt{3}\, Y_1 Y_5\right) Y_8\\
\left(Y_2 Y_4-2 Y_3^2-\sqrt{6}\, Y_1 Y_5\right) Y_6
-\left(\sqrt{2}\left( Y_2 Y_3-2 Y_5^2\right )-2 \sqrt{3}\, Y_1 Y_4\right) Y_7
 \end{array}\right)
\sim\, \mathbf{4}  \,,\\[2mm]
Y_{\mathbf{4},2}^{(10)} &= 
\left(Y_1^2+2 Y_3 Y_4+2 Y_2 Y_5\right)\\
&\quad \times
 \left(\begin{array}{c}
\sqrt{2} \left(\sqrt{3}\, Y_1 Y_6+Y_5 Y_7\right)  Y_7
+ \left(Y_3 Y_6-\sqrt{2}\, Y_4 Y_8\right) Y_8 
\\
\left(2 Y_2 Y_7+Y_4 Y_8-\sqrt{2}\, Y_3 Y_6\right) Y_6 
+ \sqrt{2}  \left(Y_3 Y_7+Y_5 Y_8\right) Y_8
\\
\left(2 Y_5 Y_8+Y_3 Y_7 -\sqrt{2}\, Y_4 Y_6\right)Y_6
+ \sqrt{2} \left(Y_4 Y_8 + Y_2 Y_7\right) Y_7
\\
\sqrt{2} \left(\sqrt{3}\, Y_1 Y_6+Y_2 Y_8\right)  Y_8
+ \left(Y_4 Y_6-\sqrt{2}\, Y_3 Y_7\right) Y_7
 \end{array}\right)
\sim\, \mathbf{4}  \,,\\[2mm]
Y_{\mathbf{4},3}^{(10)} &= 
\left(3 \sqrt{3}\, \left(Y_2 Y_3^2+Y_4^2 Y_5\right)+\sqrt{2}\, Y_1 \left(Y_1^2+3 Y_3 Y_4-6 Y_2 Y_5\right)\right)
 \left(\begin{array}{c}
\sqrt{2}\, Y_7 Y_9+Y_8 Y_{10} \\
-\sqrt{2}\, Y_6 Y_{10}-Y_8 Y_{11} \\
-\sqrt{2}\, Y_6 Y_{11}-Y_7 Y_{10} \\
\sqrt{2}\, Y_8 Y_9+Y_7 Y_{11}
 \end{array}\right)
\sim\, \mathbf{4}  \,,\\[2mm]
Y_{\mathbf{5},1}^{(10)} &= 
\left(Y_1^2+2 Y_3 Y_4+2 Y_2 Y_5\right)^2
 \left(\begin{array}{c}
 Y_1 \\ 
 Y_2 \\ 
 Y_3 \\
 Y_4 \\
 Y_5
 \end{array}\right)
\sim\, \mathbf{5}  \,,
\end{aligned} \nonumber
\end{equation}
%
\begin{equation}
\begin{aligned}
Y_{\mathbf{5},2}^{(10)} &= 
\left(Y_1^2+2 Y_3 Y_4+2 Y_2 Y_5\right)
\\ \times&
 \left(\begin{array}{c}
 2 \sqrt{2}\, Y_1 \left(Y_1^2-Y_3 Y_4-4 Y_2 Y_5\right)
-3 \sqrt{3} \left(Y_2 Y_3^2+Y_4^2 Y_5\right)
\\
Y_1 \left(\sqrt{3} \left(Y_4^2-2 Y_3 Y_5\right)-\sqrt{2}\, Y_1 Y_2\right)
+\sqrt{2} \left(3 Y_3^3+Y_2 Y_4 Y_3-2 Y_5 \left(Y_2^2-3 Y_4 Y_5\right)\right)
\\
\sqrt{2} \left(Y_2 \left(3 Y_4^2+4 Y_3 Y_5\right)-2 Y_3^2 Y_4\right)
-4 Y_1 \left(\sqrt{3}\, Y_2^2+\sqrt{2}\, Y_1 Y_3+\sqrt{3}\, Y_4 Y_5\right)
\\
\sqrt{2} \left(Y_5 \left(3 Y_3^2+4 Y_2 Y_4\right) -2 Y_3 Y_4^2\right)
-4 Y_1 \left(\sqrt{3}\, Y_5^2+\sqrt{2}\, Y_1 Y_4+\sqrt{3}\, Y_2 Y_3\right)
\\
Y_1 \left(\sqrt{3} \left(Y_3^2-2 Y_2 Y_4\right)-\sqrt{2}\, Y_1 Y_5\right)
+\sqrt{2} \left(3 Y_4^3+Y_3 Y_5 Y_4-2 Y_2 \left(Y_5^2-3 Y_2 Y_3\right)\right)
 \end{array}\right)
\sim\, \mathbf{5}  \,,\\[2mm]
Y_{\mathbf{5},3}^{(10)} &= 
 \left(\begin{array}{c}
\left(2 Y_1 \left(Y_1^2+3 Y_3 Y_4-6 Y_2 Y_5\right) + 3 \sqrt{6} \left(Y_2 Y_3^2+Y_4^2 Y_5\right)\right) \left(Y_6^2-Y_7 Y_8\right)
\\
-\left(2 \sqrt{3}\, Y_1 \left(Y_1^2+3 Y_3 Y_4-6 Y_2 Y_5\right) + 9 \sqrt{2} \left(Y_2 Y_3^2+Y_4^2 Y_5\right)\right) Y_6 Y_7
\\
\left(\sqrt{6}\, Y_1 \left(Y_1^2+3 Y_3 Y_4-6 Y_2 Y_5\right)
+ 9 \left( Y_2 Y_3^2 + Y_4^2 Y_5 \right)\right) Y_7^2
\\
\left(\sqrt{6}\, Y_1 \left(Y_1^2+3 Y_3 Y_4-6 Y_2 Y_5\right)
+ 9 \left( Y_2 Y_3^2 + Y_4^2 Y_5 \right)\right) Y_8^2
\\
-\left(2 \sqrt{3}\, Y_1 \left(Y_1^2+3 Y_3 Y_4-6 Y_2 Y_5\right) + 9 \sqrt{2} \left(Y_2 Y_3^2+Y_4^2 Y_5\right)\right) Y_6 Y_8
 \end{array}\right)
\sim\, \mathbf{5}  \,,\\[2mm]
Y_{\mathbf{5},4}^{(10)} &= 
\left(3 \sqrt{3}\, \left(Y_2 Y_3^2+Y_4^2 Y_5\right)+\sqrt{2}\, Y_1 \left(Y_1^2+3 Y_3 Y_4-6 Y_2 Y_5\right)\right)
\\ &\quad \times
 \left(\begin{array}{c}
 \sqrt{2} \left(Y_1^2+Y_3 Y_4-2 Y_2 Y_5\right)\\
 \sqrt{3} \,Y_4^2 - 2 \sqrt{2}\, Y_1 Y_2\\
 \sqrt{2} \,Y_1 Y_3 + 2 \sqrt{3}\, Y_4 Y_5\\
 \sqrt{2} \,Y_1 Y_4 + 2 \sqrt{3}\, Y_2 Y_3\\
 \sqrt{3} \,Y_3^2 - 2 \sqrt{2}\, Y_1 Y_5
 \end{array}\right)
\sim\, \mathbf{5}  \,,
\end{aligned} \nonumber
\end{equation}
%
corresponding to a total dimension of 51. 
As before, the correct dimensionality of each linear space is guaranteed via an appropriate number of constraints relating products of modular forms.

\section{Correspondence with the Dedekind Eta Function}
\label{app:Dedekind}
\begin{figure}
\centering
\includegraphics[width=10cm]{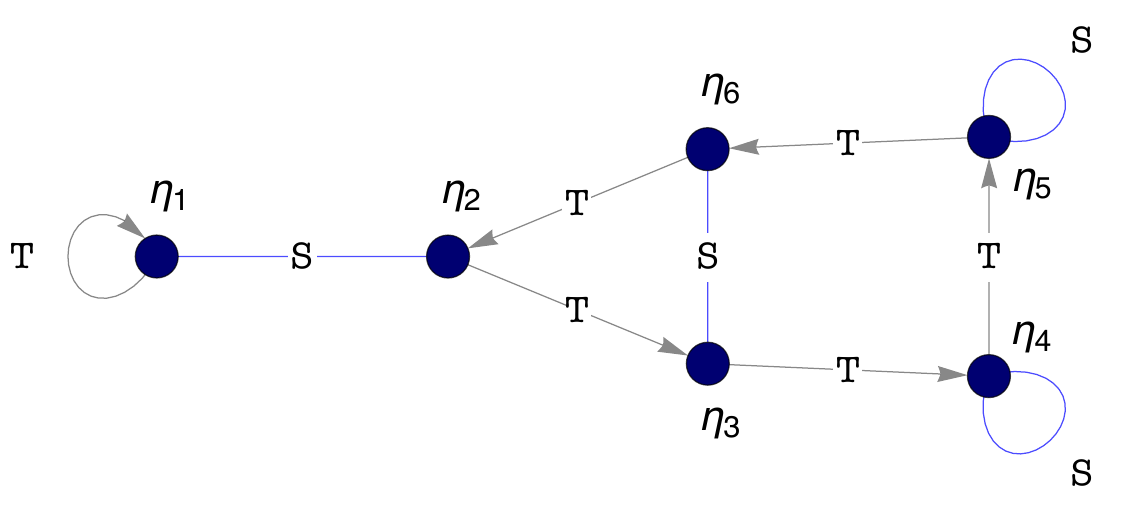}
\caption{Graph illustrating the automorphisms
of the set of seed functions $\eta_i(\tau)$,
defined in eq.~\eqref{eq:etasetA5},
under the actions of $\Gamma_5 \simeq A_5$ generators $S$ and $T$.}
\label{fig:graphEta}
\end{figure}

 For the groups $\G_2 \simeq S_3$, $\G_3 \simeq A_4$ and $\G_4 \simeq S_4$, 
the seed functions, from which the modular forms of weight 2 
are constructed, are given by the Dedekind eta function
\be
\eta(\t) \equiv q^{1/24} \prod_{n=1}^{\infty}\left(1 - q^n\right)\,,
\qquad
q = e^{2\pi i\t}\,.
\ee
%
Namely, for $\G_2$, the set of interest contains three seed functions 
$\bar\eta_i$ \cite{Kobayashi:2018vbk}: 
\be
\left\{\bar\eta_i\right\} = 
\left\{
\eta\left(2\tau\right),~
\eta\left(\frac{\tau}{2}\right),~
\eta\left(\frac{\tau+1}{2}\right)
\right\}.
\label{eq:etasetS3}
\ee
%
In the case of $\G_3$, the corresponding set of four seed functions 
$\tilde\eta_i$ reads \cite{Feruglio:2017spp}: 
\be
\left\{\tilde\eta_i\right\} = 
\left\{
\eta\left(3\tau\right),~
\eta\left(\frac{\tau}{3}\right),~
\eta\left(\frac{\tau+1}{3}\right),~
\eta\left(\frac{\tau+2}{3}\right)
\right\},
\label{eq:etasetA4}
\ee
%
while for $\G_4$, the desired set contains six seed functions $\hat\eta_i$ \cite{Penedo:2018nmg}:
\be
\left\{\hat\eta_i\right\} = 
\left\{
\eta\left(4\tau\right),~
\eta\left(\frac{\tau}{4}\right),~
\eta\left(\frac{\tau+1}{4}\right),~
\eta\left(\frac{\tau+2}{4}\right),~
\eta\left(\frac{\tau+3}{4}\right),~
\eta\left(\tau+\frac{1}{2}\right)
\right\}.
\label{eq:etasetS4}
\ee
%
It is interesting to continue this chain to the case of $\G_5 \simeq A_5$ 
considering the following set of six eta functions:
\be
\left\{\eta_i\right\} = 
\left\{
\eta\left(5\tau\right),~
\eta\left(\frac{\tau}{5}\right),~
\eta\left(\frac{\tau+1}{5}\right),~
\eta\left(\frac{\tau+2}{5}\right),~
\eta\left(\frac{\tau+3}{5}\right),~
\eta\left(\frac{\tau+4}{5}\right)
\right\}.
\label{eq:etasetA5}
\ee
%
First, we notice that this set is closed under the action of the modular group 
(and under the action of $\G_5$), i.e., 
upon acting with $S$ or $T$ 
each of the $\eta_i$ functions is mapped to another function from the set 
(up to (sometimes $\tau$-dependent) multiplicative factor).
A graph of this mapping is shown in Fig.~\ref{fig:graphEta}.
Taking logarithmic derivatives, we find:
\begin{align}
\frac{\di}{\di\tau}\log\eta_i(-1/\tau) &=
\frac{1}{2\tau}+\frac{\di}{\di\tau}\log\eta^S_i(\tau)\,,\\[2mm]
\frac{\di}{\di\tau}\log\eta_i(\tau+1) &= 
\frac{\di}{\di\tau}\log\eta^T_i(\tau)\,,
\end{align}
%
where $\eta^S_i$ and $\eta^T_i$
are the images of $\eta_i$ under the $S$ and $T$ maps of Fig.~\ref{fig:graphEta},
respectively.

 Now we construct a sum 
\be
X(a_1,\ldots,a_6|\tau) \equiv \sum_{i=1}^6 \,a_i\,\frac{\di}{\di\tau}\log \eta_i(\tau)\,,
\quad
\text{with} \quad \sum_{i} a_{i} = 0\,.
\ee
%
The functions $X$ transform as weight 2 modular forms. 
Under the action of $S$ and $T$ we find
\begin{align}
S:\,\,
X(a_1,\ldots,a_6|\tau)\,\,&\rightarrow\,\,
X(a_1,a_2,a_3,a_4,a_5,a_6|{-1/\tau})\,=\,
\tau^2 \,
X(a_2,a_1,a_6,a_4,a_5,a_3|\tau)
\,,
\\[2mm]
T:\,\,
X(a_1,\ldots,a_6|\tau)\,\,&\rightarrow\,\,
X(a_1,a_2,a_3,a_4,a_5,a_6|\tau+1)\,=\, 
X(a_1,a_6,a_2,a_3,a_4,a_5|\tau)
\,.
\end{align}
%
We search for 5 independent modular forms $X_i(\tau)$ transforming in $\5$ of $A_5$ 
according to eq.~\eqref{eq:vvmf}, which for $\g = S$ and $T$ reads
\be
X_\5\left(-1/\tau\right) = \tau^2\, \rho\left(S\right)\, X_\5\left(\tau\right)
\quad \text{and} \quad
X_\5\left(\tau+1\right) = \rho\left(T\right)\, X_\5\left(\tau\right)\,,
\ee
%
where $X_\5 = (X_1, X_2, X_3, X_4, X_5)^T$, 
and the matrices $\rho(S)$ and $\rho(T)$ 
are given by eq.~\eqref{eq:irrep5}.
The transformation $T$ fixes $a_i$ up to coefficients $c_i$, namely,
\begin{align}
X_1\left(\tau\right) &= c_1\, X\left(-5,1,1,1,1,1|\tau\right),\\
X_2\left(\tau\right) &= c_2\, X\left(0,\zeta^4,\zeta^3,\zeta^2,\zeta,1|\tau\right),\\
X_3\left(\tau\right) &= c_3\, X\left(0,\zeta^3,\zeta,\zeta^4,\zeta^2,1|\tau\right),\\
X_4\left(\tau\right) &= c_4\, X\left(0,\zeta^2,\zeta^4,\zeta,\zeta^3,1|\tau\right),\\
X_5\left(\tau\right) &= c_5\, X\left(0,\zeta,\zeta^2,\zeta^3,\zeta^4,1|\tau\right),
\end{align}
%
where $\zeta = e^{2\pi i/5}$.
The transformation $S$ fixes the coefficients $c_i$ up to an overall factor $c$:
\begin{align}
c_1 &= c\,,\\
c_2 &= -c\, \sqrt3\, \sqrt{-\varphi + i \sqrt{\sqrt5/\varphi}}\,,\qquad
c_3 = c\, \sqrt3\, \sqrt{1/\varphi - i \sqrt{\sqrt5\,\varphi}}\,,\\
c_4 &= c\, \sqrt3\, \sqrt{1/\varphi + i \sqrt{\sqrt5\,\varphi}}\,,\qquad
c_5 = -c\, \sqrt3\, \sqrt{-\varphi - i \sqrt{\sqrt5/\varphi}}\,,
\end{align}
%
with $\varphi = (1+\sqrt{5})/2$.
Choosing for convenience $c = -1/\sqrt{6}$, 
which leads to the following simplification: 
\be
c_1 = -\frac{1}{\sqrt6}\,,~~c_2 = \zeta\,,~~c_3 = \zeta^2\,,~~c_4 = \zeta^3\,,~~c_5 = \zeta^4\,,
\ee
we finally obtain
\be
X_\5(\tau) = \begin{pmatrix}
X_{1}(\tau)\\
X_{2}(\tau)\\
X_{3}(\tau)\\
X_{4}(\tau)\\
X_{5}(\tau)
\end{pmatrix}
\equiv\begin{pmatrix}
-\frac{1}{\sqrt{6}}X\left(-5,1,1,1,1,1 |\tau\right)\\
X(0,1,\zeta^4,\zeta^3,\zeta^2,\zeta|\tau)\\
X(0,1,\zeta^3,\zeta,\zeta^4,\zeta^2|\tau)\\
X(0,1,\zeta^2,\zeta^4,\zeta,\zeta^3|\tau)\\
X(0,1,\zeta,\zeta^2,\zeta^3,\zeta^4|\tau)
\end{pmatrix},
\ee
%
cf. eq.~\eqref{eq:5}. 
Thus, starting from the set of eta functions in eq.~\eqref{eq:etasetA5}, 
one can construct the $A_5$ quintet of weight 2 modular forms. 
It is interesting to notice that the graph in Fig.~\ref{fig:graphEta} 
represents a ``half'' of that in Fig.~\ref{fig:graph}, 
which in addition to the quintet allows for construction of the two $A_5$ triplets.

\bibliography{ModularA5Symmetry}

\end{document}